\documentclass[10pt,journal,twoside]{IEEEtran}
\usepackage[T1]{fontenc}
\usepackage[utf8]{inputenc}
\usepackage{amsmath,amssymb,bm,booktabs,array,graphicx,cite,microtype}
\usepackage{algorithm,algorithmic,url}
\usepackage{placeins}
\usepackage{wrapfig}
\usepackage{xcolor}
\DeclareRobustCommand{\rev}[2]{#2\label{rev:#1}}
\usepackage[hidelinks]{hyperref}
\newcommand{\E}{\mathbb E}
\newcommand{\ind}{\mathbb I}
\newcommand{\clip}{\operatorname{clip}}
\newcommand{\diag}{\operatorname{diag}}

\newcolumntype{P}[1]{>{\raggedright\arraybackslash}p{#1}}
\newtheorem{proposition}{Proposition}

\graphicspath{{figure/}}
\begin{document}
\bstctlcite{PUMAReferenceStyle}
\markboth{IEEE TRANSACTIONS ON MOBILE COMPUTING}{PUMA-Former: Uncertainty-Aware Satellite Computation Offloading}
\hypersetup{pdftitle={PUMA-Former: Uncertainty-Aware Agentic Computation Offloading in MEO-LEO Satellite Networks With Incomplete Information},pdfauthor={Yi Jing, Chunxiao Jiang, Jiawei Wang, Jiachen Sun},pdfsubject={Manuscript prepared for IEEE Transactions on Mobile Computing}}
\title{PUMA-Former: Uncertainty-Aware Agentic Computation Offloading in MEO-LEO Satellite Networks With Incomplete Information}
\author{Yi~Jing, \emph{Member, IEEE,}
Chunxiao~Jiang, \emph{Fellow, IEEE,} Jiawei~Wang, \emph{Member, IEEE,} and
Jiachen~Sun, \emph{Graduate Student Member, IEEE}
\thanks{This work was supported by the National Natural Science Foundation of China under Grant 62341131, Grant 62325108, and Grant 62501354. \textit{(Corresponding author: Chunxiao Jiang.)}}
\thanks{Y. Jing and J. Sun are with the Department of Electronic Engineering, Tsinghua University, Beijing 100084, China (e-mail: jy21@mails.tsinghua.edu.cn; eejcsun@ust.hk).}
\thanks{C. Jiang and J. Wang are with the State Key Laboratory of Space Network and Communications, Tsinghua University, Beijing 100084, China, and also with the Beijing National Research Center for Information Science and Technology, Tsinghua University, Beijing 100084, China (e-mail: jchx@tsinghua.edu.cn; w-jw@tsinghua.edu.cn).}}
\maketitle
\begin{abstract}
Satellite edge computing distributes mobile workloads across moving processors whose service capacity must be coordinated over time. In medium Earth orbit (MEO)--low Earth orbit (LEO) networks, periodic reports leave CPU states stale or missing, making offloading sensitive to uncertain capacity and shared-destination congestion. We propose PUMA-Former, an uncertainty-aware agentic framework for cooperative offloading under incomplete information that balances completed work, service delay, and execution cost. A MEO coordinator maintains distinct-report memory and closes the loop between available evidence, capacity estimates, allocation commands, and execution feedback. Latent-delay attention weights plausible measurement-age hypotheses, and historical innovations refine CPU estimates around the latest received report. Queue-error-aware fusion incorporates auxiliary pressure while accounting for backlog uncertainty. The inferred resource moments condition a structured Transformer allocator for joint destination, partition, bandwidth, CPU, and power decisions. Bounded risk-aware calibration then adjusts workload splits to reflect uncertainty and shared destination load. In traffic-driven development evaluation, PUMA-Former increases admitted compute-work completion by 8.62\%, reduces mean modeled service delay by 34.78\%, and reduces mean episode-P95 delay by 71.08\% relative to ATLA-adapt. \rev{A2}{It also improves completion and normalized energy over fixed-half round-robin offloading.} \rev{A3}{Service gains over the learning baselines involve higher normalized energy.}
\end{abstract}

\section{Introduction}
\IEEEPARstart{S}{atellite} edge computing extends mobile computing services to geographically dispersed users through low Earth orbit (LEO) constellations~\cite{jingSatelliteMEC2022,sunchinacom}. Orbital motion changes which satellites can serve a location and which neighboring processors are reachable. Inter-satellite links enable cooperative offloading beyond the access satellite, coupling communication capacity, available CPU, and queue evolution~\cite{marlSatellite,deployment}. Heterogeneous demands require allocation to consider both input volume and computation intensity~\cite{heterogeneous}. A heavily loaded satellite can benefit from neighboring processors even when aggregate constellation capacity is ample. The systems problem is to turn geographically distributed capacity into completed computational work through the currently reachable network.

In a medium Earth orbit (MEO)--LEO hierarchy, the MEO layer coordinates resources while LEO processors execute tasks. Coordination determines offloading destinations, workload partitions, bandwidth, CPU, and transmit power. Transferring more work consumes link resources and may shift congestion to a shared destination; physical capacity can also truncate a feasible command. Mobility adds a further distinction between reachability and service capability: public geometry can identify a usable neighbor without revealing how much processing capacity it currently has. Effective coordination connects both kinds of information to resource-sensitive workload assignments.

The offloading controller maintains report memory, reasons about present resources, issues executable assignments, and incorporates their consequences into subsequent decisions~\cite{agentification,agenticGnn,closedLoop}. \rev{AG1}{Here, agentic control denotes autonomous operation of this persistent perception--inference--action loop at the MEO coordinator.} \rev{AG2}{Memory and resource estimates evolve as observations arrive; deployed network parameters remain fixed.} The architectural question is how incomplete evidence should alter an executable workload assignment across shared processors.

Periodic reporting separates observation time from decision time. Delayed or lost reports leave satellites represented by measurements of different ages. Reception time need not reveal acquisition time: a newly received CPU report may describe an earlier operating condition. Delayed-state allocation~\cite{stateDelay}, robust offloading under uncertain task-computation requirements~\cite{robustCompute}, and information-age analysis~\cite{leoAge} motivate addressing this uncertainty. A reception-age feature alone cannot distinguish a recent measurement from an older one that has just arrived; padding alone cannot resolve missing and stale information.

History provides additional evidence but requires careful interpretation. Repeated cached measurements contain no new information, while averaging old resource values can conceal a recent capacity drop and encourage excessive offloading. The latest received report should provide a reference, with distinct earlier reports explaining changes in the operating regime. Unknown acquisition time further requires retaining multiple plausible delay interpretations and their implications for current capacity.

Auxiliary observations complement history but introduce another ambiguity: aggregate pressure reflects both unfinished work and processing capacity. High pressure can indicate a long queue, a slow processor, or both. Fusion must account for queue-estimation error when deciding how strongly this indirect evidence should revise CPU availability.

Graph-aware encoders and Transformer policies capture network and control dependencies~\cite{gate,atla}; optimization-informed learning preserves execution structure~\cite{optimizationDL}. Yet mean prediction alone is insufficient: destinations with similar estimated capacity can have different uncertainty and overload exposure. Exploiting this distinction requires an explicit connection from inferred resource moments to workload assignment, with bounded adjustments that preserve an effective allocation proposal.

To address these challenges, we propose PUMA-Former, an uncertainty-aware agentic computation-offloading framework. A temporal--topological encoder supplies context, while a dedicated branch infers CPU availability from distinct reports through latent-delay attention and anchored historical innovations. Queue-error-aware assimilation incorporates auxiliary pressure. The fused moments condition a structured hybrid decoder and a bounded risk-calibration operator. This design connects the interpretation of incomplete evidence to executable workload redistribution within the MEO--LEO feedback loop.

The main contributions are summarized as follows.

\begin{itemize}
\setlength{\itemsep}{2pt}
\setlength{\parskip}{0pt}
\item We formulate MEO--LEO computation offloading under incomplete information, coupling mobile reachability and resource allocation within a closed execution loop.
\item We develop anchored CPU inference through latent-delay attention, historical innovations, and queue-aware auxiliary fusion to estimate capacity and uncertainty.
\item We design a structured Transformer allocator with bounded risk-aware calibration, translating resource uncertainty into route-specific workload partitions.
\end{itemize}

The remainder of this paper is organized as follows. Section~II reviews related work. Section~III presents the system model, and Section~IV formulates the allocation problem. Section~V develops PUMA-Former and its training and execution algorithms. Section~VI reports the evaluation, and Section~VII concludes the paper.

\section{Related Work}
\subsection{Satellite Resource Allocation}
Satellite-orchestrated low-altitude resource management~\cite{jingOrchestration2026} illustrates hierarchical coordination across aerial and orbital resources. The present formulation focuses on the inference and allocation interfaces of periodic resource reports, with LEO processors executing the admitted workload.

\begin{table}[!t]
\caption{Research Positioning}
\label{tab:position}
\centering\footnotesize
\begin{tabular}{P{.24\columnwidth}P{.32\columnwidth}P{.34\columnwidth}}
\toprule
Direction & Established focus & Focus of this work\\\midrule
Delayed satellite state~\cite{stateDelay} & Delay-MDP and augmented state & Masked, age-aware state distribution and auxiliary update\\
Robust computation~\cite{robustCompute} & Uncertain task-computation requirements & Hidden available CPU and queue state\\
GATE / ATLA~\cite{gate,atla} & Structured encoders and hybrid resource decisions & Reception-delay inference and bounded offload correction\\
Contact / routing~\cite{visibility,packetRouting,nonIsl} & Visibility and diverse routing settings & Fluid communication/computation execution\\
\bottomrule
\end{tabular}
\end{table}

Satellite cooperation includes multi-timescale caching~\cite{cache}, caching-assisted offloading~\cite{dynamicCache}, multi-layer reinforcement learning~\cite{multiLayer}, and energy-oriented allocation~\cite{energySAC}. PUMA-Former focuses on inferring the current resources needed for such redistribution.

Multihop offloading and trusted cooperation introduce path-level decisions~\cite{qosMultihop,trustedHop}; our action space retains one neighbor per source and window. Two-stage game-theoretic allocation~\cite{jingTwoStage2026} coordinates resource choices in satellite-assisted low-altitude networks. Beam hopping and cross-layer interference control~\cite{beamDemand,evolutionLEO} provide complementary communication-side load balancing.

For mobile edge services, reliability-aware scheduling~\cite{qoeReliability} and resource adaptation for deterministic satellite--terrestrial networking~\cite{deterministic} link allocation to the service experienced by workloads. PUMA-Former studies the complementary information problem: reachable processors are known, but their current service capacity is not. The focus is therefore coordinated processing under uncertain resource states, with routing scope and physical execution held common across the compared methods.

\rev{RW1}{Joint beam-pattern and power allocation~\cite{zhangBeam2026} uses multi-agent actor--critic learning to adapt communication resources to demand.} \rev{RW2}{Multi-modal offloading~\cite{gongMultimodal2026} couples task scheduling with processing and authentication, while space-based zero-trust architectures~\cite{chennetwork} address direct-to-cell security.} \rev{RW3}{These studies motivate joint resource control; PUMA-Former develops the report-age inference and uncertainty-to-partition interface needed when processor capacity is hidden.}

\subsection{Imperfect Resource Information}
Delayed-state offloading~\cite{stateDelay} and imperfect-information games~\cite{imperfectGame} address incomplete decision information. Distributionally robust on-orbit scheduling~\cite{robustOrbit} accounts for uncertain remote-sensing data volumes. PUMA-Former represents latent acquisition-delay alternatives and propagates their disagreement into CPU uncertainty used for allocation.

Freshness-oriented relay and cyclic-update scheduling~\cite{relayAge,cyclicAge} addresses information value over time. Here, observed reception age and hidden acquisition age are separated without exposing simulator timestamps.

CNN--LSTM resource forecasting~\cite{jingForecast2026} models spatiotemporal demand. Rhythmic resource sensing~\cite{jingRhythmic2026} studies state acquisition, while RhyTrust-Sat~\cite{jingRhyTrust2026} uses signal evidence to guide that acquisition. PUMA-Former takes arrived reports as its information interface and uses their age and content to infer present processing capacity for offloading.

\begin{figure}[!t]
\centering\includegraphics[width=\columnwidth,trim=0bp 8bp 0bp 8bp,clip]{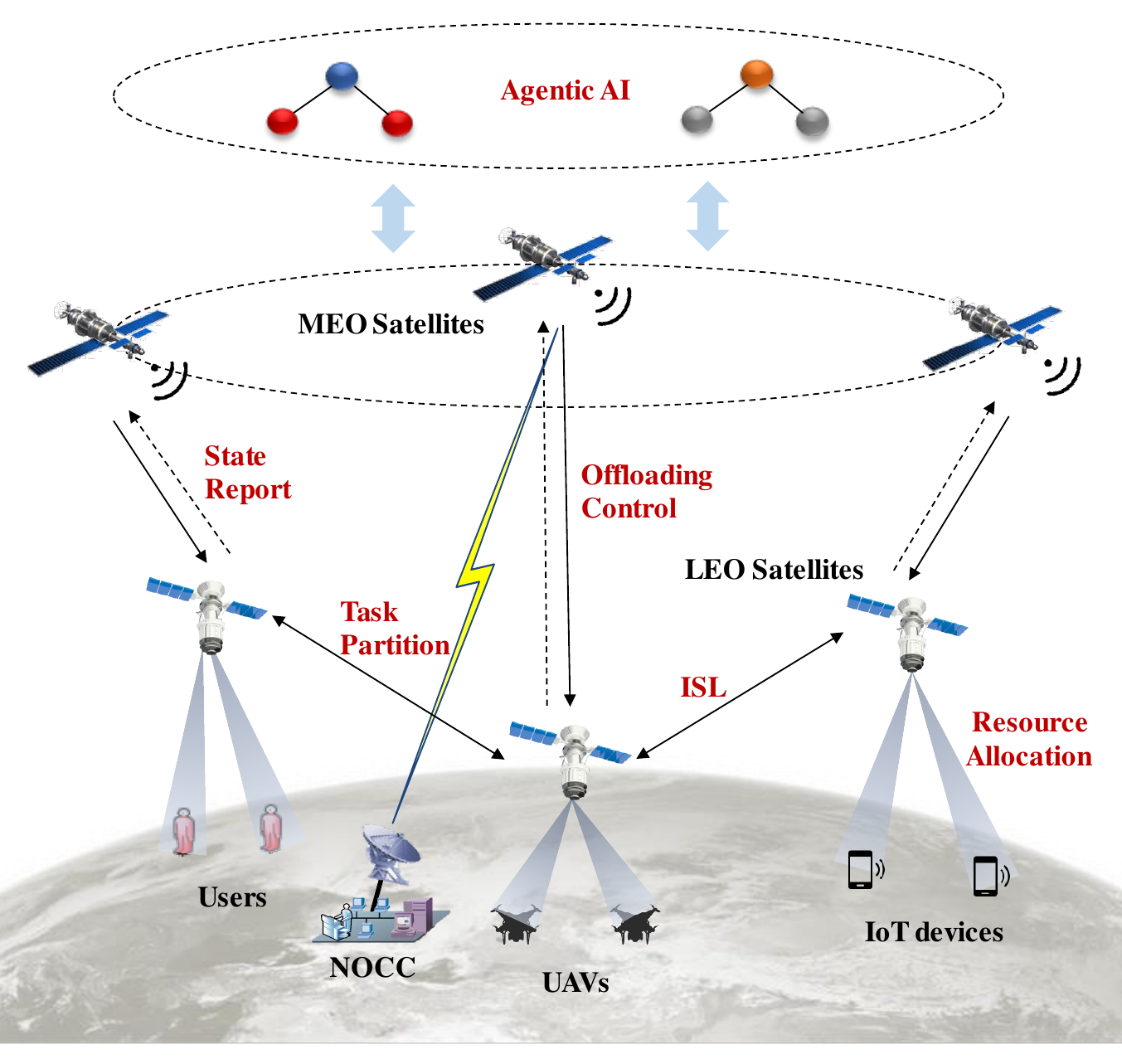}
\caption{Hierarchical satellite computation offloading. The MEO coordination layer receives LEO state reports and issues offloading controls; LEO satellites partition admitted tasks and exchange workload over ISLs. The NOCC supports management, while user, UAV, and IoT icons represent heterogeneous access-demand sources. Task processing in the evaluated model takes place at the LEO layer.}
\label{fig:system}
\end{figure}

\subsection{Structured Network Intelligence}
Graph-structured allocation~\cite{graphSagin}, temporal graph encoding~\cite{gate}, and Transformer policies~\cite{atla,dagTransformer} capture entity and temporal dependencies. LLM-guided control~\cite{llmSatellite}, diffusion-based collaboration~\cite{lowAltitude}, and agentic retrieval~\cite{agenticRetrieval} explore broader reasoning mechanisms. PUMA-Former specializes these architectural principles to report memory, CPU inference, and bounded execution.

Digital-twin-enabled scheduling in OCEAN~\cite{jingOcean2026} provides a related MEO--LEO coordination setting. PUMA-Former develops report-age inference and bounded workload adjustment for LEO processors with incompletely observed capacity.

Table~\ref{tab:position} summarizes these distinctions. PUMA-Former connects report identity and hidden acquisition age to CPU moments and offload calibration.

\section{System Model}
\label{sec:system}
Fig.~\ref{fig:system} presents the hierarchical offloading system. This section defines its execution rules and observation model.
\subsection{Network Architecture}
Consider $N$ LEO satellites indexed by $\mathcal N$, coordinated by a logical agentic controller in the MEO layer. Resource reports travel upward to this controller, and allocation commands return to LEO nodes. The network operations and control center (NOCC) supports network management. Users, IoT devices, and UAV terminals in Fig.~\ref{fig:system} illustrate possible demand sources; the evaluated workload represents their access traffic through a common timestamp--location--volume interface. Public orbital geometry defines visible access satellites and a bounded neighbor set $\mathcal N_i(t)$, consistent with visibility-dependent satellite networking~\cite{visibility,packetRouting,nonIsl}. A Walker constellation supplies the simulated geometry. Time is divided into demand windows of duration $\Delta$, with queues attached to physical satellite identities across windows.

Satellite service chaining~\cite{sdnSpace} and cloud-native operating systems~\cite{11730502} provide resource-orchestration mechanisms. In our model, the MEO layer and NOCC coordinate tasks that execute on LEO processors. Access bandwidth is aggregated per satellite, distinct from detailed cooperative beam scheduling~\cite{cooperativeBeam}.

The model separates mobility from resource uncertainty. Geometry updates the access association and $\mathcal N_i(t)$ at decision windows; these public quantities define the admissible destinations. CPU availability and queue occupancy are observed through separate reporting channels. A neighbor can thus remain reachable while its reported capacity becomes stale. Conversely, a previously useful report does not make a currently unreachable node an admissible destination. Candidate masks enforce the current topology, whereas resource inference determines how the reachable candidates are used.

\rev{T1}{The window length $\Delta$ is an allocation and demand-aggregation interval, distinct from individual task duration.} Effective link rates and persistent satellite queues couple geometry changes to inter-window backlog.

MEO coordination provides a common view of arrived reports and competing offload requests, while LEO nodes enforce local transmission and CPU limits. A logical coordinator is assumed over the modeled domain. Report and command delivery form the control interface; handover among coordinators is outside the evaluated allocation model. Table~\ref{tab:notation} summarizes the principal symbols.

\begin{table}[t]
\caption{Principal Symbols}
\label{tab:notation}\centering\footnotesize
\begin{tabular}{P{.28\columnwidth}P{.64\columnwidth}}
\toprule Symbol & Meaning\\\midrule
$\Delta,\mathcal N_i(t)$ & Demand-window length and reachable neighbors\\
$B_i,C_i,Q_i$ & New input bits, new CPU cycles, retained cycles\\
$\omega_i,F_0,F_i$ & Available fraction, nominal and executed CPU rate\\
$\mathbf a_i,\mathbf a_i',\bar a_i$ & Proposed command, calibrated command, realized offload fraction\\
$j_i,a_i,b_i,f_i,p_i$ & Destination, partition, bandwidth, CPU, power controls\\
$X_i,U_i,Z_i$ & Transferred bits, executed cycles, dropped cycles\\
$d_{ir}^{\mathrm{rx}},\delta_{ir}$ & Reception age and hidden report delay in windows\\
$\mu_i,\sigma_i;\mu_i^+,\sigma_i^+$ & CPU moments before and after auxiliary fusion\\
$y_i,\tau_q,\epsilon_a$ & Pressure, retained-work budget, correction limit\\
$\mathcal I_t,\mathcal M_t$ & Available information and distinct-report memory\\
\bottomrule
\end{tabular}
\end{table}

\subsection{Traffic Mapping}
Satellite-assisted edge applications motivate mapping communication requests into processing demands~\cite{ocean,jing2026multigranularitycommunicationcomputingresourcemapping}; heterogeneous-task studies distinguish input volume from compute intensity~\cite{heteroTasks}. For traffic record $\ell$, the input contains a UTC timestamp, latitude, longitude, and volume $V_\ell$ in decimal MB. Its average communication rate and mapped computation are
\begin{equation}
r_\ell=\frac{8\times10^6V_\ell}{\Delta},\quad
B_\ell=8\times10^6\eta V_\ell,\quad C_\ell=\kappa_\ell B_\ell ,
\label{eq:map}
\end{equation}
where $\eta$ is the computing-traffic fraction and $\kappa_\ell$ is a synthetic cycles/bit intensity. Thus, measured traffic drives the workload, while application CPU demand is model-derived. Visible records are associated with LEO satellites using a common access procedure. Summing the mapped workload gives input bits $B_i(t)$ and cycles $C_i(t)$. Uncovered demand contributes to the coverage denominator; admitted demand enters the satellite computing system. The evaluation preserves the trace's original geographic coverage.

The representative input size $b_0$ converts each aggregate into job count $n_i(t)$ for fluid delay accounting. A 300-s window contains many second-scale jobs: $\kappa_\ell$ determines CPU load, whereas $b_0$ determines service granularity at that load.

Common settings ensure identical offered work and representative-job granularity across policies.

\subsection{Resource Execution}
The command supplied to the physical executor is
\begin{equation}
\mathbf a_i(t)=(j_i, a_i,b_i,f_i,p_i),\qquad
j_i\in\{i\}\cup\mathcal N_i(t),
\label{eq:action}
\end{equation}
Here $\mathbf a_i$ is the action vector and $a_i$ the offload fraction. The other controls specify destination, bandwidth, CPU, and power. Each source selects at most one remote neighbor; multiple sources may share a destination.

Section~V distinguishes proposal $\mathbf a_i$ from calibrated command $\mathbf a_i'$; the executor then applies the following saturation limits.

Let $\omega_i(t)\in[\omega_{\min},1]$ denote available CPU fraction and $F_0$ nominal cycles/s. The executed CPU share is
\begin{equation}
s_i=f_{\min}+(1-f_{\min})f_i,\qquad F_i=s_iF_0\omega_i.
\label{eq:cpu}
\end{equation}
Availability follows a bounded exogenous process with autoregressive, diurnal, and hotspot components. Distinguishing nominal from available CPU captures time-varying processing conditions relevant to dynamic satellite computing~\cite{thermal}.

For a selected ISL, the effective rate is
\begin{equation}
R_i=R_0 b_i g_{ij_i}
\frac{\log_2(1+9p_i)}{\log_2(10)},
\label{eq:rate}
\end{equation}
where $g_{ij_i}$ is the bounded geometry factor. Requested and executed fractions differ when link capacity is insufficient:
\begin{equation}
X_i=\min\{ a_i B_i,R_i\Delta\},\qquad
\bar a_i=X_i/\max(B_i,\epsilon).
\label{eq:transmit}
\end{equation}
For local execution, $X_i=\bar a_i=0$. Untransferred input remains local. Equation~\eqref{eq:rate} assigns each source an aggregate-link budget; shared destination CPU creates the inter-source processing contention.

Let $Q_i(t)$ be unfinished CPU cycles. The processing demand and executed work are
\begin{align}
L_i&=Q_i+(1-\bar a_i)C_i+
       \sum_{k:j_k=i,\ k\ne i}\bar a_kC_k,\\
U_i&=\min\{L_i,F_i\Delta\},\qquad
Z_i=[L_i-U_i-F_i\tau_q]^+,\label{eq:work}\\
Q_i(t+1)&=\min\{[L_i-U_i]^+,F_i\tau_q\}.
\label{eq:queue}
\end{align}
Here $\tau_q$ is the retained-backlog waiting budget and $Z_i$ is dropped work. The finite queue records overload explicitly through both retained and dropped cycles.

\begin{proposition}[Work conservation and bounded correction]
The executed partition conserves new mapped CPU work. If an offload correction satisfies $| a_i'- a_i|\leq\epsilon_a$ and physical truncation uses~\eqref{eq:transmit}, then the change in transmitted input is at most $\epsilon_a B_i$ for fixed destination and rate.
\end{proposition}
\begin{IEEEproof}
Summing local and remote contributions yields
$\sum_i(1-\bar a_i)C_i+\sum_i\bar a_iC_i=\sum_iC_i$.
The function $x\mapsto\min\{x,R_i\Delta\}$ is one-Lipschitz. Applying it to $ a_iB_i$ and $ a_i'B_i$ gives the stated single-step perturbation bound.
\end{IEEEproof}

Shared destination capacity $F_i\Delta$ couples otherwise individually feasible source actions through~\eqref{eq:work}--\eqref{eq:queue}.

\subsection{Service Accounting}
\label{sec:service}
Communication, queueing, and computation jointly determine offloading latency, a coupling also present in edge scheduling and parallel execution~\cite{edgeSlam,parallelMec}. Our delay estimator combines representative-job service with bounded queueing. If $\bar c_i$ is mean job work at processor $i$, define
\begin{align}
t_i^{\mathrm{svc}}&=\bar c_i/F_i,\qquad
\widetilde\rho_i=\min\{\rho_i,0.98\},\\
\tau_i^{\mathrm{wait}}&=\min\left\{\min\{Q_i/F_i,\tau_q\}
+\frac{t_i^{\mathrm{svc}}\widetilde\rho_i}
{2(1-\widetilde\rho_i)},\tau_q\right\}.
\label{eq:wait}
\end{align}
Here $\rho_i$ is newly assigned work divided by slot CPU capacity. \rev{T2}{Demand within a window is represented by an average offered rate and a representative job size.} The second term is an M/D/1-inspired approximation; clipping $\rho_i$ at 0.98 regularizes it near saturation. A similarly bounded expression models ISL waiting. Excess work is accounted for by the execution and drop equations. This approximation estimates representative service under the allocated rates, including finite backlog pressure.

For a source with nonempty local and remote branches,
\begin{equation}
d_i=t_i^{\mathrm{acc}}+
\max\{t_i^{\mathrm{svc}}+\tau_i^{\mathrm{wait}},\ 
t_i^{\mathrm{isl}}+t_{ij_i}^{\mathrm{prop}}+
t_{j_i}^{\mathrm{svc}}+\tau_{j_i}^{\mathrm{wait}}\}.
\label{eq:delay}
\end{equation}
An absent branch is excluded from the maximum. \rev{T3}{Both branches follow access service; remote processing follows ISL transfer and propagation.} Split processing completes with the slower branch in~\eqref{eq:delay}. Window-level transferred and processed volumes use average rates over $\Delta$, while the delay expression applies the service precedence to representative jobs.

\rev{T4}{Retained work contributes $Q_i/F_i$ to the next window's backlog term, capped by $\tau_q$.} The state tracks unfinished cycles without individual arrival timestamps, so this term represents carried processing pressure. \rev{T5}{Calendar waiting between windows is outside this delay approximation.} \rev{T6}{The 30-s value is a reference threshold for modeled service risk.} \rev{T7}{Return-link delivery, contact waiting, control-message delivery, and coordinator computation are also outside~\eqref{eq:delay}.} Reported completion and modeled delay jointly describe the allocation operating point. In particular, the window capacity determines executed work, whereas $b_0$ and the effective rates determine representative-job delay. \rev{T8}{These quantities support allocation comparisons at a common granularity; task-level deadline compliance requires arrival-time-resolved service accounting.}

With processor utilization $u_i=U_i/(F_i\Delta)$, energy is
\begin{equation}
\begin{aligned}
E_t=\sum_i\big[&\Delta(P_{\mathrm{idle}}\ind\{L_i>0\}
+P_{\mathrm{dyn}}s_i^3u_i)\\
&+p_iP_{\max}X_i/R_i\big].
\end{aligned}
\label{eq:energy}
\end{equation}
The transmission term is zero for a local action. The resulting energy metric combines modeled processor activity and offloading transmission.

\subsection{Observation Model}
\label{sec:observation}
Satellite reports are generated periodically with satellite-dependent phases and may arrive late or be lost. For report $r$, let $t_{ir}^{\mathrm{rx}}$ and $g_{ir}$ be its reception and generation slots. Only the reception age
\begin{equation}
d_{ir}^{\mathrm{rx}}(t)=t-t_{ir}^{\mathrm{rx}}
\label{eq:age}
\end{equation}
is supplied to the policy. Acquisition age is $d_{ir}^{\mathrm{rx}}+\delta_{ir}$, where $\delta_{ir}=t_{ir}^{\mathrm{rx}}-g_{ir}$ is hidden. A report labeled Fresh can therefore contain an old measurement.

Observations contain resource values, masks, reception age, and Fresh/Stale/Missing indicators. Zero-filled entries retain validity masks. A length-$L$ history de-duplicates copies with the same inferred reception stamp. A delayed old report may replace a newer measurement, so reception order is not assumed to equal generation order.

\rev{R1}{Reports are due every six windows with a fixed random phase per satellite.} \rev{R2}{At each scheduled report, loss is an independent Bernoulli event with probability 0.05.} \rev{R3}{Conditional on delivery, delay is zero with probability 0.70 and is uniform over one, two, and three windows with total probability 0.30.} \rev{R4}{The inference alternatives $\{0,1,2,3\}$ use these window units.} \rev{R5}{The evaluated training and development runs share this channel configuration.} \rev{R6}{The controller interface exposes reception time; acquisition timestamps remain hidden.} \rev{R7}{Such ambiguity can arise when an aggregation interface exports resource values and reception metadata without verifiable sample-time metadata.} The model addresses this information interface; the available reception age and latent acquisition delay are kept separate throughout inference.

An additional synthetic sensor observes neighborhood pressure:
\begin{equation}
y_i=\clip\left(\sum_jG_{ij}
 \frac{q_j+w_j^{\mathrm{arr}}}{\omega_j}+\epsilon_i,\ 0,\ 10\right),
\label{eq:sensor}
\end{equation}
where $G$ is the row-normalized neighbor matrix and $\epsilon_i$ is zero-mean Gaussian noise with standard deviation 0.08. The sensor uses $q_j=\clip(Q_j/(F_0\tau_q),0,10)$ and $w_j^{\mathrm{arr}}=\clip(C_j/(F_0\Delta),0,10)$. Their sum is a dimensionless pressure feature with separate queue and arrival normalizations. The controller receives the noisy scalar $y_i$ and its validity; predicted queue moments and current demand descriptors support its interpretation.

\rev{Y1}{Equation~\eqref{eq:sensor} specifies an additional current-window observation channel.} A possible implementation summarizes locally measured backlog and processing-rate indicators and forwards an aggregate to the coordinator. \rev{Y2}{The evaluation synthesizes this measurement from physical states and assumes it is available at the decision instant.} Its freshness follows this channel assumption. Sensor acquisition, aggregation, and delivery delays are not resolved in the modeled task-service budget. Table~\ref{tab:information} identifies the information available at deployment.

The corresponding deployment interface is a lightweight indicator service alongside periodic resource reporting. \rev{Y3}{A local monitor forms the normalized workload-to-processing indicator, and aggregation over the known neighbor weights produces $y_i$.} \rev{Y4}{The actor receives the noisy aggregate and validity flag; raw current queues and CPU fractions remain hidden.} A reserved indicator channel supplies current-window evidence under the stated observation model.

Methods using auxiliary observations share~\eqref{eq:sensor}; clipping, aggregation, and queue error limit its information.

\section{Problem Formulation}
We seek low-delay completion of admitted work with controlled execution cost under incomplete information. Reliability-aware and differentiated scheduling~\cite{qoeReliability,deterministic,statisticalQos} motivate the joint service criterion.
\subsection{Information State}
\label{sec:information}
\begin{figure}[!t]
\centering\includegraphics[width=\columnwidth]{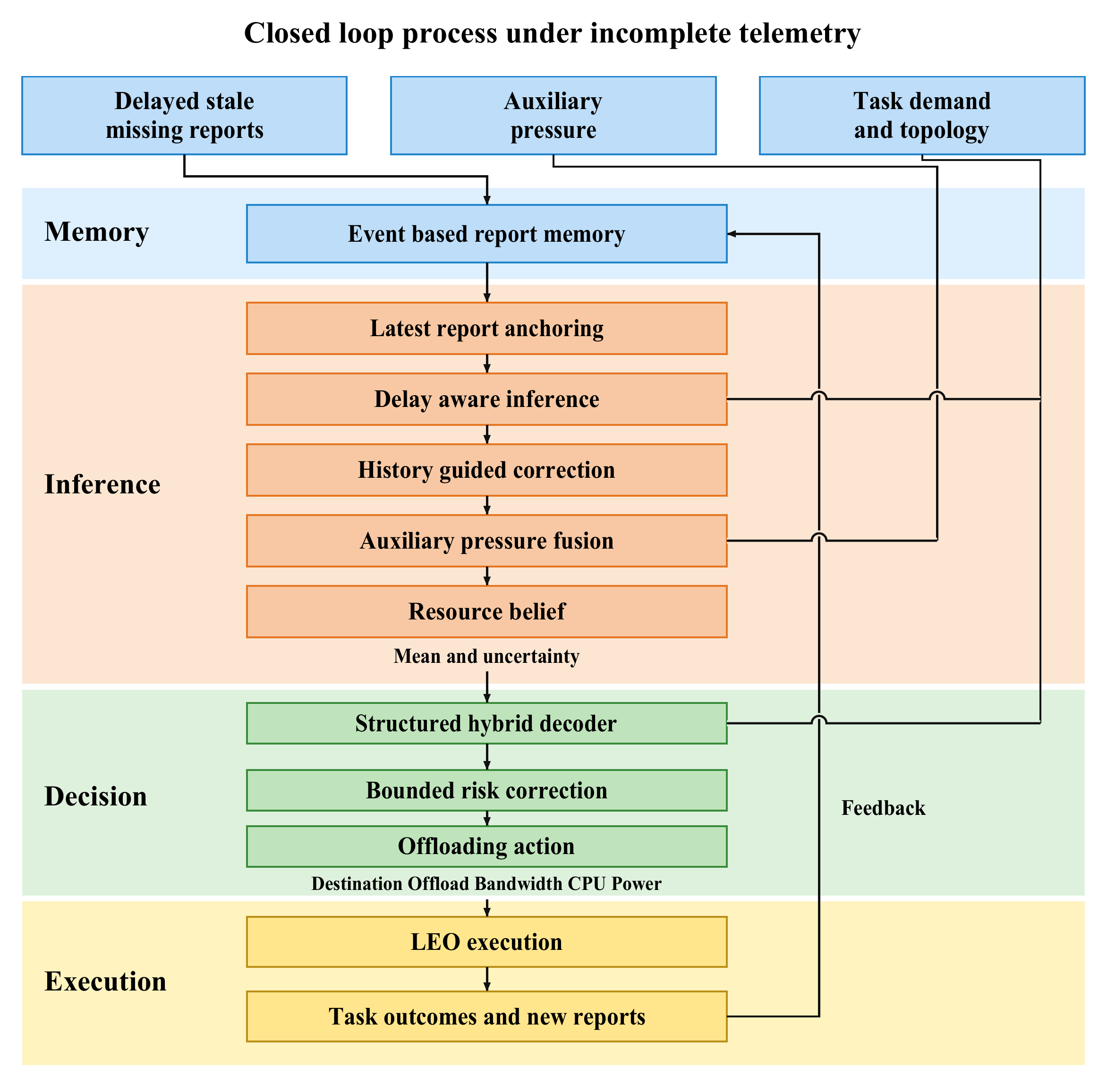}
\caption{Closed-loop offloading under incomplete information. Distinct reports update event memory, while auxiliary pressure directly enters resource fusion. Task demand and topology condition inference and allocation. LEO execution changes queues and subsequent reports, providing feedback for the next decision window.}
\label{fig:loop}
\end{figure}

Fig.~\ref{fig:loop} specifies interaction across decision windows. \rev{I1}{The information set $\mathcal I_t$ contains arrived reports, report memory, auxiliary pressure and its validity, current demand descriptors, and public topology.} \rev{I2}{Current CPU and queue states remain latent; true CPU values serve only as training labels.}

\begin{table}[t]
\caption{Information Access}
\label{tab:information}\centering\footnotesize
\begin{tabular}{P{.41\columnwidth}P{.20\columnwidth}P{.29\columnwidth}}
\toprule Quantity & Online access & Role\\\midrule
Arrived resource values & Yes; possibly stale & Reports and memory\\
Reception age, masks, status & Yes & Evidence interpretation\\
Demand descriptors, topology & Yes & Tasks and candidate masks\\
Noisy pressure, validity & Yes; channel assumption & Auxiliary fusion\\
Current true CPU and queue & Hidden & Execution and training labels\\
Report generation time, true delay & Hidden & Channel simulation\\
Predicted CPU and queue moments & Computed online & Decoding and calibration\\
\bottomrule
\end{tabular}
\end{table}

At the start of window $t$, arrived reports $O_t$ update the event memory:
\begin{equation}
\mathcal M_t=\mathcal U_L(\mathcal M_{t-1},O_t),
\label{eq:memoryupdate}
\end{equation}
where $\mathcal U_L$ removes duplicate reception stamps, retains masks, and enforces horizon $L$. \rev{I3}{Only reports enter memory; pressure enters assimilation, while task and topology supply context.} An admissible policy $\pi_\theta(\mathbf a_t\mid\mathcal I_t)$ respects this information boundary.

Let $S_t$ collect physical queues, available CPU, link conditions, and current demand. The physical transition is determined by the execution rules and exogenous arrivals, whereas the observation channel determines which components reach the controller. A compact description is
\begin{equation}
S_{t+1}\sim P(\cdot\mid S_t,\mathbf a_t),\qquad
O_{t+1}\sim P_{\mathrm{obs}}(\cdot\mid S_{0:t+1}).
\label{eq:partial}
\end{equation}
Received reports can describe earlier states, so the observation law depends on history. Commands change backlog and pressure through~\eqref{eq:transmit}--\eqref{eq:queue}; arrived observations then convey those effects to the controller. Section~V specifies the mapping from $\mathcal I_t$ to commands. Table~\ref{tab:information} applies to all deployed learned policies; the full-information reference has explicitly privileged access to current resource states.

\subsection{Allocation Objective}

Let $\Pi_{\mathcal I}$ be the policies measurable with respect to $\mathcal I_t$: they cannot condition an action on hidden current CPU or an unreceived report. The allocation problem is
\begin{subequations}\label{eq:objective}
\begin{align}
\mathrm{P1}:\quad\max_{\pi\in\Pi_{\mathcal I}}\quad
&\E_{\pi}\!\left[\sum_{t=0}^{T-1}\gamma^t r_t\right],\label{eq:utility}\\
\text{s.t.}\quad &j_i(t)\in\{i\}\cup\mathcal N_i(t),\label{eq:routeconstraint}\\
&\mathbf a_t\in\mathcal A_t,\quad F_i=s_iF_0\omega_i,\label{eq:boxconstraint}\\
&0\leq X_i\leq\min\{ a_i B_i,R_i\Delta\},\label{eq:linkconstraint}\\
&0\leq U_i\leq\min\{L_i,F_i\Delta\},\label{eq:computeconstraint}\\
&Q_i(t+1)=[L_i-U_i-Z_i]^+,\label{eq:queueconstraint}\\
&0\leq Q_i(t+1)\leq F_i\tau_q,\quad\forall i,t.\label{eq:backlogconstraint}
\end{align}
\end{subequations}
Equations~\eqref{eq:transmit}--\eqref{eq:queue} determine $X_i$, $U_i$, and $Z_i$ from the control action and available resources. The executor attains the limits in~\eqref{eq:linkconstraint} and~\eqref{eq:computeconstraint}. Constraints~\eqref{eq:routeconstraint} and~\eqref{eq:boxconstraint} enforce reachable destinations and control ranges; the remaining constraints express link capacity, processing capacity, and finite backlog. Resource limits can therefore reduce the executed portion of a feasible command.
The implemented reward is
\begin{align}
r_t={}&4c_t-0.25\phi(\bar d_t/10)-0.03e_t
-0.35\phi(\bar w_t/8)\nonumber\\
&-(0.30+\lambda_d)v_t-(0.70+\lambda_o)o_t
-g_t+0.70\psi_t .
\label{eq:reward}
\end{align}
Here $c_t=\sum_iU_i/\sum_i(Q_i+C_i)$ is slot work completion,
$\bar d_t$ and $\bar w_t$ are representative-job-weighted mean service and waiting times in seconds over active sources. Their reference values are 10 and 8~s, respectively. Further, $e_t$ is energy normalized by the network slot budget,
$v_t$ is the job-weighted maximum of deadline exceedance and attributed drop fraction, $o_t$ combines link and queue overload, $g_t$ is requested-versus-executed offload mismatch, and $\psi_t$ is clipped reduction in peak normalized processing load. The bounded cost is
$\phi(x)=\log(1+\min\{x,8\})/\log 9$.
Projected updates of the nonnegative multipliers adapt the penalties to observed risk and overload. These terms provide soft service control; hard probabilistic guarantees require explicit constraint treatment, as illustrated by peak-age-constrained scheduling~\cite{safeAge}.

Completion rewards processing, while delay, service risk, energy, overload, and execution mismatch penalize its costs. Peak-load reduction encourages redistribution; the coefficients specify the service--energy preference.

The hybrid feasible set can be written as
\begin{equation}
\begin{aligned}
\mathcal A_t=\{\mathbf a:\;&j_i\in\{i\}\cup\mathcal N_i(t),\quad 0\leq f_i\leq1,\\
& ( a_i,b_i,p_i)\in\mathcal B\ \text{if }j_i\ne i,\\
&X_i=0\ \text{if }j_i=i,\quad \forall i\},
\end{aligned}
\label{eq:feasible}
\end{equation}
where $\mathcal B=[ a_{\min},1]\times[b_{\min},1]\times[p_{\min},1]$ is the remote-control box. Link truncation and queue service enforce execution limits; destination contention and hidden capacity remain coupled through the transition and reward.

P1 couples discrete destinations, shared queues, and hidden capacity. Motivated by goal-directed collaboration~\cite{goalCollaborator} and learned--analytical offloading~\cite{diffusionOffload}, PUMA-Former combines resource inference, a hybrid allocator, and bounded workload calibration to approximate this partially observable control problem.

The policy optimizes destination and resource commands; physical execution determines $X_i$ and $U_i$. Two sources may each select a feasible destination while jointly exceeding its CPU capacity. The transition and reward capture the resulting processing, backlog, and drops. P1 consequently evaluates a sequence of allocations: work left by an earlier decision affects later service even if future arrivals and topology are unchanged.

\subsection{Evaluation Metrics}
\label{sec:metrics}

Three evaluation quantities are distinguished:
\begin{align}
\mathrm{CR}_{\mathrm{comp}}
 &=\min\left\{1,\frac{\sum_{t,i}U_i(t)}{\sum_{t,i}C_i(t)}\right\},\label{eq:cr}\\
\mathrm{CR}_{\mathrm{proxy}}
 &=\mathrm{CR}_{\mathrm{comp}}\,v_{\mathrm{cov}},\label{eq:proxy}\\
\mathcal E_{\mathrm{norm}}
 &=10^9\,\frac{\sum_tE_t}{\sum_{t,i}U_i(t)}.
\label{eq:eff}
\end{align}
\rev{M3}{Coverage $v_{\mathrm{cov}}$ is served communication volume divided by original offered volume.} \rev{M4}{Equation~\eqref{eq:proxy} defines a coverage-adjusted proxy using this volume ratio and admitted compute-work completion.} Its denominator combines two workload measures; individual original-demand task delivery would require task-level accounting. Energy in~\eqref{eq:eff} is measured in joules per completed Gcycle.

\rev{D1}{Within each episode, mean delay and P95 weight representative source jobs by their counts.} \rev{D2}{The episode P95 is the weighted 95th percentile of the modeled delays.} \rev{D3}{Across episodes, the reported tail measure averages these episode P95 values.} \rev{D4}{Mean delay describes the workload's typical modeled service cost, while the tail measure identifies less favorable service conditions.}

Global utilization uses network CPU capacity; active utilization includes only satellites with demand.

\rev{M1}{The completion denominator counts each newly admitted cycle once.} With an empty initial queue, summing~\eqref{eq:work}--\eqref{eq:queue} gives
\begin{equation}
\sum_{t,i} C_i(t)=\sum_{t,i}U_i(t)+\sum_{t,i}Z_i(t)+\sum_iQ_i(T).
\label{eq:episodebalance}
\end{equation}
This identity separates executed, dropped, and terminal unfinished work. The slot reward instead normalizes against all work available for service in that slot, including backlog. Hence training return measures sequential decision utility, while episode completion measures the final disposition of newly admitted work.

\section{Proposed Scheme}
\label{sec:scheme}
Fig.~\ref{fig:network} details the within-window resource inference and action generation of the interaction loop in Fig.~\ref{fig:loop}. A context encoder represents tasks, topology, and report history; an anchored branch estimates current CPU availability. Its mean and uncertainty summarize the resource needed for allocation. Denote fixed backbone parameters by $\theta$ and trainable report-inference parameters by $\varphi$. The resulting computation is
\begin{equation}
\begin{aligned}
(\mu_t^+,\sigma_t^+)&=\mathcal G_{\theta,\varphi}(\mathcal I_t),\\
\mathbf a_t&\sim\pi_\theta(\cdot\mid\mathcal I_t,\mu_t^+,\sigma_t^+),\\
\mathbf a_t'&=\mathcal C(\mathbf a_t,\mu_t^+,\sigma_t^+;\boldsymbol\ell_t,\mathbf w_t,\widehat{\mathbf s}_t).
\end{aligned}
\label{eq:interface}
\end{equation}
Here $\mathcal G_{\theta,\varphi}$ includes inference and assimilation. Calibration uses pre-transfer loads $\boldsymbol\ell_t$, new work $\mathbf w_t$, and route-weighted shares $\widehat{\mathbf s}_t$ defined in Section~V-E. All inputs derive from $\mathcal I_t$. Execution~\eqref{eq:transmit}--\eqref{eq:queue} uses $\mathbf a_t'$, producing the realized fraction $\bar a_i=\min\{a_i'B_i,R_i\Delta\}/\max\{B_i,\epsilon\}$.

Online evaluation of~\eqref{eq:interface} uses the most probable destination and transformed conditional means.

\subsection{Policy Architecture}
\begin{figure*}[!t]
\centering\includegraphics[width=\textwidth]{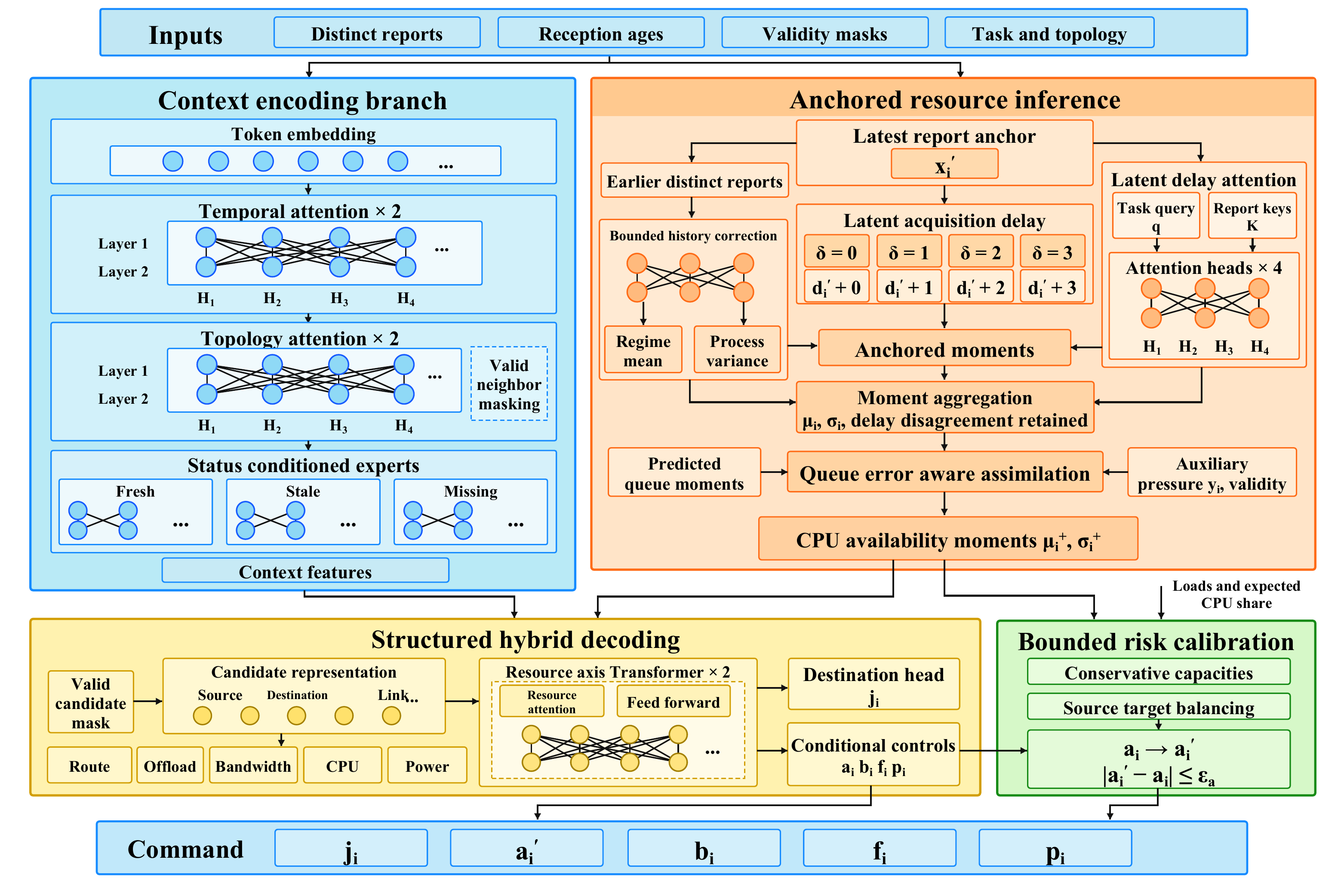}
\caption{PUMA-Former inference and allocation architecture. Temporal and topology attention produce context features. Latest-report anchoring, latent-delay attention, and historical correction estimate CPU availability; queue-aware assimilation fuses auxiliary pressure. The resulting moments support structured hybrid decoding and bounded risk calibration, yielding $(j_i,a_i',b_i,f_i,p_i)$.}
\label{fig:network}
\end{figure*}

The context encoder embeds reports, ages, masks, and tasks. Two temporal layers precede two topology layers with neighbor masking. Multi-head attention, residual connections, normalization, and feed-forward transformations use width 128, four heads, and feed-forward width 256. Softmax weights combine three experts using encoded features and age/status metadata. The Fresh/Stale/Missing labels in Fig.~\ref{fig:network} identify conditioning regimes; a learned status prior biases the mixture rather than imposing hard routing.

Let $h_i$ denote the context encoder output. Its reconstruction and scale heads supply the normalized queue mean $\widehat q_i$ and uncertainty $\sigma_i^q$:
\begin{equation}
\widehat q_i=10\,\operatorname{sigmoid}(z_i^q-4),\qquad
\sigma_i^q=0.01+0.60\,\operatorname{sigmoid}(z_i^s),
\label{eq:queueheads}
\end{equation}
where $z_i^q,z_i^s$ are projections of $h_i$. These heads implement the ``Predicted queue moments'' input in Fig.~\ref{fig:network} using context features, not an external queue measurement. They exclude raw pressure and remain fixed during CPU-inference fitting.

Report-based CPU inference can run alongside context encoding. It combines the latest received anchor, earlier-report innovations, and four delay hypotheses. Auxiliary assimilation starts after both CPU and queue moments are available. Raw pressure enters only this operator. Its outputs $(\mu_i^+,\sigma_i^+)$ summarize the dimensionless availability $\omega_i$; physical processing rate is $F_i=s_iF_0\omega_i$.

The decoder combines context and queue/CPU moments into source--destination--edge tokens. Five embeddings represent route, offload, bandwidth, CPU, and power. Two resource-axis Transformer layers produce destination logits and route-conditioned Gaussian parameters. Squashing enforces minimum remote offload, bandwidth, power, and executed CPU share of 0.15, 0.50, 0.50, and 0.65, respectively.

The set $\mathcal C_i=\{i\}\cup\mathcal N_i$ includes local processing and reachable neighbors, whose spare capacity remains uncertain. With inferred moments implicit in $\mathcal I_t$, the policy factorizes as
\begin{equation}
\pi_i(\mathbf a_i\mid\mathcal I_t)=
\pi_i^{\mathrm{route}}(j_i\mid\mathcal I_t)
\pi_i^{\mathrm{cont}}( a_i,b_i,f_i,p_i\mid j_i,\mathcal I_t).
\label{eq:factorization}
\end{equation}
Calibration changes only $a_i$ for fixed decoder outputs; inferred moments can also affect all controls through decoding.

\subsection{Latent Delay Attention}
\label{sec:latent}
For each distinct usable report $r$, construct delay alternatives
$\delta\in\mathcal D=\{0,1,2,3\}$ and effective age
\begin{equation}
d_{ir\delta}=d_{ir}^{\mathrm{rx}}+\delta.
\label{eq:latentage}
\end{equation}
The reporting-model prior is $(0.70,0.10,0.10,0.10)$. An eight-feature task descriptor $\mathbf t_i$ and a six-feature report key $\mathbf z_{ir\delta}$, containing CPU, effective age, delay hypothesis, phase descriptors, and queue, are projected into four heads of width 16:
\begin{equation}
e_{ir\delta}^{h}
=\frac{(W_Q^h\mathbf t_i)^\top W_K^h\mathbf z_{ir\delta}}{\sqrt{16}}.
\label{eq:att}
\end{equation}
Padding and unavailable reports are masked after de-duplication. Even at zero reception age, the alternatives span acquisition ages zero through three; they age together until another report arrives.

\subsection{Anchored History Inference}
\label{sec:history}
An $8$--$32$--$2$ task-conditioned network with a tanh hidden layer outputs $(u_i^{\mathrm{pr}},v_i^{\mathrm{pr}})$ and defines
\begin{equation}
\begin{aligned}
m_i&=\omega_{\min}+(1-\omega_{\min})\operatorname{sigmoid}(u_i^{\mathrm{pr}}),\\
\sigma_{i,\mathrm{prior}}&=0.01+\operatorname{softplus}(v_i^{\mathrm{pr}}).
\end{aligned}
\label{eq:prior}
\end{equation}
Its parameters belong to $\varphi$. The mean stays in $(\omega_{\min},1)$ and the scale is positive. Without a usable CPU report, $(\mu_i,\sigma_i)=(m_i,\sigma_{i,\mathrm{prior}})$ in both training and deployment.

Otherwise, let $r^\star$ be the most recently received distinct usable report. Following Fig.~\ref{fig:network}, its value and reception age are $x_i'=x_{ir^\star}$ and $d_i'=d_{ir^\star}^{\mathrm{rx}}$. The prime marks the received anchor, not a newly generated measurement. For head $h$ with mean-reversion coefficient $\rho_h\in(0,1)$, an older report has gap $b_{ir}=d_{ir}^{\mathrm{rx}}-d_i'\geq0$ and innovation
\begin{equation}
I_{ir}^{h}=
\frac{x_i'-m_i-\rho_h^{b_{ir}}(x_{ir}-m_i)}
{\max(1-\rho_h^{b_{ir}},0.10)}.
\label{eq:innovation}
\end{equation}
Attention over earlier reports produces a bounded trend and surprise:
\begin{align}
T_i^h&=\sum_{r\ne r^\star}\beta_{ir}^h\tanh(I_{ir}^h/0.20),\\
S_i^h&=\sum_{r\ne r^\star}\beta_{ir}^h
        \min\{(I_{ir}^h/0.20)^2,4\}.
\label{eq:history}
\end{align}
For the valid older-report set $\mathcal R_i^-$, delay marginalization and normalization are
\begin{equation}
z_{ir}^h=\log\sum_{\delta\in\mathcal D}p_\delta e^{e_{ir\delta}^h},\qquad
\beta_{ir}^h=\frac{e^{z_{ir}^h}}{\sum_{s\in\mathcal R_i^-}e^{z_{is}^h}}.
\label{eq:historyweights}
\end{equation}
If $\mathcal R_i^-$ is empty, both summaries are zero.

Mean reversion toward $m_i+\Delta m$ predicts $m_i+\rho_h^b(x_{ir}-m_i)+(1-\rho_h^b)\Delta m$; solving for $\Delta m$ gives~\eqref{eq:innovation}. The floor and bounded transformation control inconsistent evidence. Learned gains regulate the use of reception gaps as approximate temporal separation.

Historical innovations adjust the regime while preserving the latest-report anchor:
\begin{align}
m_i^h&=\clip(m_i+0.20\tanh(\xi_h)T_i^h,\omega_{\min},1),\\
\mu_{i\delta}^h&=m_i^h+\rho_h^{d_i'+\delta}(x_i'-m_i^h),\\
v_{i\delta}^h&=
\left(v_{\mathrm{obs},h}
+v_{\mathrm{proc},h}\frac{1-\rho_h^{2(d_i'+\delta)}}{1-\rho_h^2}\right)
\exp\{\tanh(\zeta_h)S_i^h\}.
\label{eq:conditional}
\end{align}
History gains initialize at zero, preserving anchor-only behavior initially. Softplus transformations ensure positive noise parameters.

\rev{W1}{For example, at window 10 a report received at window 9 has reception age 1.} \rev{W2}{Delay alternatives $\delta=0,1,2,3$ assign effective ages 1, 2, 3, and 4.} An earlier distinct report received at window 7 has reception gap 2. \rev{W3}{For illustrative values $m_i=0.70$, $\rho_h=0.90$, $x_{ir}=0.60$, and $x_i'=0.65$, the uncorrected model predicts a latest value of 0.62.} \rev{W4}{Equation~\eqref{eq:innovation} gives $I_{ir}^h=0.16$, indicating an upward departure from that prediction; a positive historical gain raises the inferred reversion level.} \rev{W5}{These illustrative values explain the operator independently of fitted parameters.} Repeated cached copies share one reception identity and supply no additional event.

Delay weights include the prior and a variance term:
\begin{equation}
\lambda_{i\delta}^h\ \propto\
\exp\{e_{ir^\star\delta}^h+\log p_\delta-\tfrac12\log v_{i\delta}^h\}.
\label{eq:delayweight}
\end{equation}
Normalizing delay weights within each head yields
\begin{align}
\mu_i&=\frac14\sum_{h,\delta}\lambda_{i\delta}^h\mu_{i\delta}^h,\\
\sigma_i^2&=\frac14\sum_{h,\delta}\lambda_{i\delta}^h
\left[v_{i\delta}^h+(\mu_{i\delta}^h-\mu_i)^2\right].
\label{eq:totalvariance}
\end{align}
The variance retains conditional variability and disagreement about report age. Zero effective acquisition age recovers the anchor; zero reception age alone does not. Windows without new reports still use retained history.

Conditional variance describes uncertainty within one head and delay interpretation, whereas the squared mean deviation retains disagreement among interpretations. Historical corrections adjust the reversion regime and variance around the received anchor, preserving the latest measurement while explaining operating changes.

\subsection{Auxiliary Assimilation}
The pressure model is approximately linear in inverse CPU, $\chi_i=1/\omega_i$. Using predicted normalized queue $\widehat q_i$, set
\begin{equation}
\begin{aligned}
A_{ij}&=G_{ij}(\widehat q_j+w_j^{\mathrm{arr}}),\qquad \widehat y=A\widehat\chi,\\
D&=\diag\left((\sigma_i/\mu_i^2)^2\right).
\end{aligned}
\label{eq:linear}
\end{equation}
Sensor noise includes both nominal measurement noise and uncertainty in the queue contribution:
\begin{equation}
R_{ii}=\sigma_\epsilon^2+
\sum_jG_{ij}^2(\sigma_j^q/\mu_j)^2+\nu_{\mathrm{mis}},
\label{eq:noise}
\end{equation}
where $\nu_{\mathrm{mis}}=\operatorname{softplus}(\zeta_{\mathrm{mis}})>0$ is fitted during backbone preparation and then fixed. Set $\widehat\chi_i=1/\mu_i$. This center and $D$ use a first-order inverse transformation, not exact inverse-resource moments. Diagonal $R$ neglects cross-observation queue-error correlations caused by overlapping neighborhoods. After excluding unavailable or endpoint-clipped pressure rows, the update is
\begin{align}
\widehat\chi^+&=\widehat\chi+
DA^\top(ADA^\top+R)^{-1}(y-A\widehat\chi),\label{eq:assimilation}\\
D^+&=D-DA^\top(ADA^\top+R)^{-1}AD.
\label{eq:fusedvariance}
\end{align}
A linear solve implements the update. With $\widetilde\chi_i^+=\clip(\widehat\chi_i^+,1,1/\omega_{\min})$, the returned mean is $\mu_i^+=1/\widetilde\chi_i^+$ and its uncertainty is $(\mu_i^+)^2\sqrt{D_{ii}^+}$, clipped to $[0.01,1]$. The queue moments are unchanged. Raw pressure enters exclusively through this operator; its influence depends on queue and sensor uncertainty.

With positive diagonal $D$ and $R$, the inverse-CPU update minimizes the local weighted least-squares objective
\begin{equation}
\frac12\|\chi-\widehat\chi\|_{D^{-1}}^2+
\frac12\|y-A\chi\|_{R^{-1}}^2,
\label{eq:fusionobjective}
\end{equation}
where $\|z\|_M^2=z^\top Mz$. The stationary condition gives~\eqref{eq:assimilation}; subsequent physical clipping bounds the deployed update.

Unavailable channels leave the prior unchanged.

Through~\eqref{eq:noise}, uncertain backlog increases observation noise and reduces the update gain. Smaller entries of $D$ likewise protect precise report-based estimates. Fusion jointly weights resource confidence and observation ambiguity.

\rev{U6}{The returned standard deviation is a local uncertainty summary after linearization, inverse transformation, and clipping.} It quantifies the model's resource ambiguity for allocation. The report-based and post-assimilation moments have different statistical meanings: the former are fitted with a Gaussian scoring rule, while the latter propagate that uncertainty through a local approximation. Their decision value and the available pre-assimilation coverage diagnostics are evaluated separately in Section~VI.

\subsection{Risk-Aware Action Calibration}
Let $\mu_i^+$ and $\sigma_i^+$ be fused CPU moments. A conservative normalized capacity is
\begin{equation}
\widehat c_i=\max\{\mu_i^+-\sigma_i^+,\omega_{\min}\}\,\widehat s_i,
\label{eq:riskcapacity}
\end{equation}
For destination probabilities $\pi_{ij}^{\mathrm{route}}$, let $\bar f_{ij}$ be the CPU control obtained by squashing the conditional Gaussian mean. The route-weighted share is
\begin{equation}
\widehat s_i=f_{\min}+(1-f_{\min})
\sum_{j\in\mathcal C_i}\pi_{ij}^{\mathrm{route}}\bar f_{ij}.
\label{eq:expectedshare}
\end{equation}
This averages over routes, not over Gaussian control samples. It provides a common capacity estimate for candidate comparisons; actual execution uses the selected $f_i$ in~\eqref{eq:cpu}. Define
\begin{equation}
w_i=\clip(C_i/(F_0\Delta),0,10),\qquad
\ell_i=w_i+\widehat q_i\tau_q/\Delta.
\label{eq:referenceload}
\end{equation}
These are pre-transfer loads: local arrivals plus estimated carried backlog, without incoming work from other sources' current commands. The factor $\tau_q/\Delta$ converts the queue feature to slot-capacity units. Balancing source and target utilization yields
\begin{equation}
 a_{ij}^{\mathrm{bal}}=
\frac{\ell_i\widehat c_j-\ell_j\widehat c_i}
{\max\{w_i(\widehat c_i+\widehat c_j),\epsilon\}}.
\label{eq:balance}
\end{equation}
The route-conditioned fraction is corrected by
\begin{equation}
 a'_{ij}=\Pi_{[ a_{\min},1]}
\left[ a_{ij}+\epsilon_a
\tanh\left(\frac{ a_{ij}^{\mathrm{bal}}- a_{ij}}{0.10}\right)\right].
\label{eq:correction}
\end{equation}
The implementation calibrates transformed conditional means for all active remote candidates, leaving destination logits unchanged. The selected route then gives $ a_i= a_{ij_i}$ and $ a_i'= a_{ij_i}'$. At deterministic evaluation, calibrating all candidates and then selecting $j_i$ is equivalent to selecting $j_i$ first and calibrating that candidate. Fig.~\ref{fig:network} depicts the selected-command interface. Local routes and sources without new work bypass correction.

To derive~\eqref{eq:balance}, consider transferring normalized workload $ a w_i$ from source $i$ to candidate $j$. The corresponding local load estimates are $\ell_i- a w_i$ and $\ell_j+ a w_i$. Equating their capacity-normalized values gives
\begin{equation}
\frac{\ell_i- a w_i}{\widehat c_i}
=\frac{\ell_j+ a w_i}{\widehat c_j},
\label{eq:equalization}
\end{equation}
whose solution is~\eqref{eq:balance}. Each transfer is evaluated against pre-transfer loads. The bounded step limits each source's deviation from its proposal; joint destination contention remains governed by~\eqref{eq:work}--\eqref{eq:queue}.

For $w_i>0$ and inactive floors, hold all decoder outputs, fused means, and reference loads fixed. This also fixes $\widehat s_i$. The direct calibration pathway then satisfies
\begin{align}
\frac{\partial a_{ij}^{\mathrm{bal}}}{\partial\sigma_j^+}
&=-\frac{\widehat s_j\widehat c_i(\ell_i+\ell_j)}
{w_i(\widehat c_i+\widehat c_j)^2}\leq0,\label{eq:targetrisk}\\
\frac{\partial a_{ij}^{\mathrm{bal}}}{\partial\sigma_i^+}
&=\frac{\widehat s_i\widehat c_j(\ell_i+\ell_j)}
{w_i(\widehat c_i+\widehat c_j)^2}\geq0.\label{eq:sourcerisk}
\end{align}
Monotone calibration preserves these directions until saturation: destination uncertainty discourages incoming work, whereas source uncertainty favors redistribution. These are local partial derivatives, not total sensitivities of the decoder or guarantees of joint service.

\subsection{Learning Protocol}
\label{sec:learning}
The backbone is prepared from a fixed PPO-trained teacher. On training-only observations, distillation matches its destination probabilities and route-conditioned Gaussian means before squashing. For teacher probabilities $\pi_i^{\mathrm T}$ and latent means $\mathbf u_{ij}^{\mathrm T}$, the loss is
\begin{equation}
\mathcal L_{\mathrm{dist}}=\mathbb E_{\mathrm{dec}}
\left[\operatorname{KL}(\pi_i^{\mathrm T}\Vert\pi_i^{\mathrm{route}})
+\sum_j\pi_{ij}^{\mathrm T}\|\mathbf u_{ij}-\mathbf u_{ij}^{\mathrm T}\|_{M_j}^2\right],
\label{eq:distillation}
\end{equation}
where $\mathbb E_{\mathrm{dec}}$ averages decision-active sources and $\|\cdot\|_{M_j}^2$ averages squared error over executed controls: CPU only locally and all four controls remotely.

Preparation fits backbone reconstruction heads to current queue and CPU labels $\mathbf z_i=(q_i,\omega_i)$; the anchored branch subsequently replaces the CPU estimate. Let $\mathbf m_i^0,\mathbf m_i^+$ denote pre-/post-assimilation estimates and $\boldsymbol\sigma_i^0$ the pre-assimilation scales. The reconstruction loss is
\begin{equation}
\mathcal L_{\mathrm{rec}}=\mathbb E_{\mathrm{status}}
\left[\|\mathbf m_i^+-\mathbf z_i\|_{\mathrm{av}}^2
+0.25\|\mathbf m_i^0-\mathbf z_i\|_{\mathrm{av}}^2
+0.05\mathcal G_i\right],
\label{eq:preparation}
\end{equation}
where $\|\cdot\|_{\mathrm{av}}^2$ averages the two components, $\mathcal G_i$ is their Gaussian NLL using detached mean errors, and $\mathbb E_{\mathrm{status}}$ weights represented report statuses equally. The preparation objective is $\mathcal L_{\mathrm{dist}}+0.05\mathcal L_{\mathrm{rec}}$. Reconstruction uses detached context features; its post-assimilation error trains $\nu_{\mathrm{mis}}$ through the fusion solve. The teacher is used only in preparation.

This prepared backbone, queue heads, and mismatch parameter are then fixed as $\theta$; only $\varphi$ is adapted. Four training dates provide 1152 fitting observations per seed, and two further training dates provide 576 check observations. Current CPU labels never enter the deployed policy.

The inference loss uses report-based CPU moments \emph{before} auxiliary assimilation:
\begin{equation}
\mathcal L=
\frac{\sum_{t,i}\varpi_{ti}\left[
(\mu_{ti}-\omega_{ti})^2/(2\sigma_{ti}^2)+\log\sigma_{ti}\right]}
{\sum_{t,i}\varpi_{ti}},
\label{eq:loss}
\end{equation}
where half the weight is uniform over satellites and half is normalized over decision-active nodes. Training uses 12 fixed epochs, batch size 16, Adam learning rate $3\times10^{-4}$, and gradient norm clipping at one. \rev{U7}{The final epoch is used without validation checkpoint selection.} Negative Gaussian NLL is valid for continuous densities.

Equation~\eqref{eq:loss} omits the Gaussian constant $\tfrac12\log(2\pi)$, consistently across variants. \rev{I4}{Ground-truth CPU and queue values supervise preparation and offline diagnostics; deployment computes resource moments from the information in Table~\ref{tab:information}.} The teacher supplies a policy initialization through distillation. The retained service comparison evaluates the resulting complete controller and its specified inference interventions.

Correction strength is selected from $\{0,0.01,0.03\}$ on fixed training-only episodes. A positive strength must preserve completion and modeled mean/P95 delay relative to zero correction, allowing a $10^{-6}$ numerical tolerance; accepted candidates are ranked by training return. The reported matched ablations use the same strength, 0.03, to avoid confounding architecture with independently selected correction magnitudes.

Algorithm~\ref{alg:train} separates inference fitting from training-only action acceptance and exports fixed parameters for Algorithm~\ref{alg:control}.

\begin{algorithm}[t]
\caption{PUMA-Former Inference Training and Calibration}
\label{alg:train}
\begin{algorithmic}[1]
\REQUIRE Fixed backbone $\theta$; training and check sets $\mathcal D_{\mathrm{fit}},\mathcal D_{\mathrm{check}}$; epochs $E$; step candidates $\mathcal S$.
\ENSURE Inference parameters $\varphi$ and accepted step $\epsilon_a$.
\STATE \textbf{Stage 1: Resource inference}
\STATE Initialize $\varphi$; set historical innovation gains to zero.
\FOR{$e=1,\ldots,E$}
  \FOR{each minibatch $\mathcal B\subset\mathcal D_{\mathrm{fit}}$}
    \STATE Form report/delay masks; compute the task prior~\eqref{eq:prior} and attention scores~\eqref{eq:att}.
    \STATE Compute report moments by~\eqref{eq:innovation}--\eqref{eq:totalvariance}; if no usable report exists, use~\eqref{eq:prior}.
    \STATE Evaluate $\mathcal L(\varphi;\mathcal B)$ in~\eqref{eq:loss} using training CPU labels.
    \STATE Update $\varphi$ with Adam; clip gradient norm at one.
  \ENDFOR
\ENDFOR
\STATE \textbf{Stage 2: Action acceptance}
\STATE On $\mathcal D_{\mathrm{check}}$, run zero correction to obtain $(c_0,d_0,d_{95,0})$; set $\mathcal S_{\mathrm{ok}}=\{0\}$.
\FOR{each $s\in\mathcal S\setminus\{0\}$}
  \STATE Execute Algorithm~\ref{alg:control} with $\epsilon_a=s$ on the same training-check episodes.
  \IF{$c_s\geq c_0$, $d_s\leq d_0$, and $d_{95,s}\leq d_{95,0}$}
    \STATE Add $s$ to $\mathcal S_{\mathrm{ok}}$.
  \ENDIF
\ENDFOR
\STATE Choose $\epsilon_a\in\mathcal S_{\mathrm{ok}}$ by training-check return; export $(\theta,\varphi,\epsilon_a)$.
\end{algorithmic}
\end{algorithm}

\subsection{Online Inference}
Algorithm~\ref{alg:control} orders the operators in Fig.~\ref{fig:network}: queue and CPU inference precede assimilation, decoding, and calibration. A missing usable CPU report activates the task prior; unavailable pressure bypasses assimilation; local routes bypass remote calibration. Fixed parameters and evolving memory implement the feedback loop in Fig.~\ref{fig:loop}.

\begin{algorithm}[t]
\caption{PUMA-Former Online Computation Offloading}
\label{alg:control}
\begin{algorithmic}[1]
\REQUIRE Fixed $(\theta,\varphi,\epsilon_a)$; public topology; report-memory length $L$.
\ENSURE Executed commands $\{\mathbf a_i'(t)\}$ and carried report memory.
\STATE Initialize masked report memory $\mathcal M_0$.
\FOR{each decision window $t$}
  \STATE \textbf{Perception and memory}
  \STATE Receive $O_t$ and valid auxiliary $y_t$; update $\mathcal M_t$ from $O_t$ using~\eqref{eq:memoryupdate}.
  \STATE Encode context; compute queue moments~\eqref{eq:queueheads} and task priors~\eqref{eq:prior}.
  \FOR{each satellite $i\in\mathcal N$}
    \IF{a usable CPU report remains in memory}
      \STATE Identify $r^\star$; compute delay ages and attention scores via~\eqref{eq:latentage}--\eqref{eq:att}.
      \STATE Compute innovations and $(\mu_i,\sigma_i)$ via~\eqref{eq:innovation}--\eqref{eq:totalvariance}.
    \ELSE
      \STATE Use $(m_i,\sigma_{i,\mathrm{prior}})$ from~\eqref{eq:prior}.
    \ENDIF
  \ENDFOR
  \STATE \textbf{Resource reasoning}
  \IF{valid auxiliary measurements are available}
    \STATE Form $A,D,R$; apply~\eqref{eq:assimilation}--\eqref{eq:fusedvariance} and inverse transformation to obtain $(\mu^+,\sigma^+)$.
  \ELSE
    \STATE Set $(\mu^+,\sigma^+)=(\mu,\sigma)$.
  \ENDIF
  \STATE \textbf{Coordination and execution}
  \STATE Decode route probabilities and conditional control means via~\eqref{eq:factorization}.
  \STATE Form $\widehat s,w,\ell$ via~\eqref{eq:expectedshare}--\eqref{eq:referenceload} and capacities via~\eqref{eq:riskcapacity}.
  \FOR{each active source $i$ and reachable remote candidate $j$}
    \STATE Compute $ a_{ij}^{\mathrm{bal}}$ and $ a_{ij}'$ via~\eqref{eq:balance}--\eqref{eq:correction}.
  \ENDFOR
  \STATE Select $j_i$ and its controls; submit $\mathbf a_i'=(j_i, a_{ij_i}',b_i,f_i,p_i)$, with local bypass.
  \STATE Apply~\eqref{eq:transmit}--\eqref{eq:queue} to $\mathbf a_i'$ to obtain $X_i,U_i,Z_i,Q_i(t+1)$.
  \STATE Retain memory; await the next arrived reports.
\ENDFOR
\end{algorithmic}
\end{algorithm}

\subsection{Execution Cost}
\label{sec:cost}
\rev{Q1}{With $L$ reports, $D_\delta$ delay alternatives, and fixed head width, the anchored inference branch requires $O(NLD_\delta)$ time and storage.} The context backbone adds temporal self-attention with quadratic interaction cost in the history length and dense topology attention with quadratic interaction cost in the node count. Neighborhood masks constrain information exchange within these attention operators.

Decoding uses five resource tokens per candidate. If $K_{\max}$ bounds the number of remote candidates, route-conditioned decoding and calibration process at most $N(K_{\max}+1)$ candidate groups. \rev{Q2}{For $M$ valid pressure observations, assimilation solves an $M\times M$ system; the dense worst case $M=N$ costs $O(N^3)$ time and $O(N^2)$ memory.} A linear-system solve implements~\eqref{eq:assimilation} without explicitly forming the inverse.

\rev{Q3}{The MEO coordinator maintains memory and executes inference; LEO nodes enforce local service limits.} Windows without new reports reuse evidence with updated ages, and unavailable auxiliary rows are excluded from fusion. The rate in~\eqref{eq:rate} accounts for task-data transfer; control-message transmission and platform-specific inference time lie outside this modeled service budget.

If each auxiliary update carries $b_y$ payload bits per satellite every $T_y$ seconds, its aggregate scalar payload rate is $Nb_y/T_y$. Packet headers, measurement metadata, aggregation exchanges, and retransmissions add to this rate. A current-window auxiliary channel assumes acquisition and delivery complete before allocation. This payload accounting specifies the interface cost; the reported execution energy covers the terms in~\eqref{eq:energy}. \rev{Q4}{The complexity bounds characterize computational growth, while wall-clock inference time and peak memory depend on implementation and hardware.}

\rev{Y5}{For a payload illustration, one 16-bit indicator per satellite per 300-s window gives 3.84~bit/s over 72 satellites, before headers and aggregation.} If every neighbor contribution requires a separate exchange, at most $NK_{\max}b_y/T_y$ additional scalar payload bits per second are needed for a direct aggregation design. \rev{Y6}{These rates describe payload only; sensing, synchronization, channel access, and delivery determine the actual implementation overhead.} Analytical complexity and payload size delimit the deployment requirements of the evaluated coordinator.

\section{Experimental Results}
\label{sec:experiments}
This section evaluates workload completion, service delay, and execution energy under incomplete resource information. Matched inference interventions and action diagnostics examine how resource estimation affects offloading.

\subsection{Simulation Setup}
\label{sec:setup}
\rev{S1}{The simulation uses real operational traffic records from China Satcom for March--April 2025, including UTC timestamps, user locations, and traffic volumes in decimal MB.} Their original geographic concentration and temporal traffic variation are retained. \rev{S2}{Equation~\eqref{eq:map} converts these measured inputs into communication demand and modeled computation: half of the traffic is assigned a computation intensity of 500--1500 cycles/bit.} \rev{S3}{A 0.40-MB representative input defines job granularity for service-delay accounting.} These application assumptions determine CPU demand from the traffic records.

\rev{S4}{The modeled network contains 72 Walker-constellation satellites in 12 planes at 550~km altitude and $53^\circ$ inclination.} Nominal CPU is 50~Gcycles/s per node, with available fractions between 0.35 and 1; access and ISL bandwidths are 160 and 80~MHz. Each full-day episode contains 288 demand windows of 300~s. Periodic reports follow the loss and delay distribution in Section~\ref{sec:observation}; current-window noisy pressure supplements the received reports. Table~\ref{tab:settings} lists the orbit, resource, channel, and learning settings.

\begin{table}[t]
\caption{Simulation Parameters}
\label{tab:settings}\centering\footnotesize
\begin{tabular}{P{.49\columnwidth}P{.43\columnwidth}}
\toprule Parameter & Value\\\midrule
Orbit planes / satellites per plane & 12 / 6\\
Altitude / inclination & 550 km / $53^\circ$\\
Minimum elevation / ISL neighbors & $25^\circ$ / at most 4\\
Nominal orbital period & 5730 s\\
Demand window / episode length & 300 s / 288 windows\\
Nominal CPU / available fraction & 50 Gcycles/s / 0.35--1\\
CPU mean / AR coefficient & 0.70 / 0.92\\
Access / ISL bandwidth & 160 / 80 MHz\\
Nominal access / ISL rate & 320 / 160 Mbit/s\\
Maximum transmit power & 10 W\\
Compute traffic fraction & 50\%\\
Compute intensity & 500--1500 cycles/bit\\
Representative job input & 0.40 MB\\
Service-risk threshold / retained-work budget & 30 / 15 s\\
Report period / late probability & 6 windows / 0.30\\
Report loss / latent late delay & 0.05 / 1--3 windows\\
Auxiliary noise standard deviation & 0.08\\
History / backbone width / heads & 12 / 128 / 4\\
Inference epochs / minibatch & 12 / 16\\
Training seeds / environment seeds & 21, 42 / 51001, 52001\\
\bottomrule
\end{tabular}
\end{table}

\subsection{Evaluation Protocol}
\label{sec:protocol}
Training data end on April 10. \rev{P1}{Development evaluation uses April 11--14, two environment seeds, and two training seeds: 16 day episodes per learned method and eight per traditional rule.} Dates and scenarios are shared across methods. These development dates are distinct from the training-only inference fitting and action acceptance in Section~\ref{sec:learning}.

Completion is first calculated from each episode's executed and admitted CPU-cycle totals. \rev{P2}{Reported method values average episode metrics, and relative improvements compare these method means.} Delay means use representative-job weights within each episode; the tail value averages episode-level P95s. \rev{P9}{Normalized energy divides summed episode energy by summed completed work; the stored mean window energy is multiplied by 288 before aggregation.} \rev{P8}{Fig.~\ref{fig:completion} reports standard deviation across two policy-seed means; Fig.~\ref{fig:latency} shows date means.} Ablation summaries, training trajectories, and separately recorded diagnostics retain their respective aggregation levels in Figs.~\ref{fig:ablation}--\ref{fig:mechanism}.

Paired comparisons use identical dates and environmental scenarios. \rev{P3}{For each date, differences are averaged over the two scenarios and, for learned policies, the two matched training seeds.} \rev{P6}{A deterministic rule contributes once per scenario, and averaging its differences over policy seeds adds no independent rule runs.} Let $D_d$ be a date's mean PUMA-minus-reference difference. \rev{P4}{Table~\ref{tab:paired} reports $\overline D$ and the two-sided 95\% interval $\overline D\pm3.18s_D/\sqrt{4}$, using a Student-$t$ reference with three degrees of freedom.} \rev{P5}{This date-block calculation conditions on the retained seeds and assumes independent, approximately normal date differences.} The four consecutive development dates provide a limited estimate of temporal variability; the intervals concern this comparison protocol.

\subsection{Baselines}
\label{sec:baselines}
The comparison includes fixed strategies, observation-based heuristics, learned policies, and an information-advantaged reference:
\begin{itemize}
\setlength{\itemsep}{2pt}
\setlength{\parskip}{0pt}
\item \textbf{Local-only:} retains all admitted work at its access satellite and requests the full local CPU share.
\item \textbf{Fixed-half RR:} offloads half of the arriving workload to a feasible neighbor selected by round-robin rotation and requests full resource shares.
\item \textbf{Random feasible:} samples an admissible destination and continuous resource controls, with realized service limited by the common executor.
\item \textbf{Stale-greedy:} ranks neighbors using reported queue-plus-arrival pressure normalized by available CPU, with a link-rate preference, and offloads when the selected neighbor has lower estimated pressure.
\item \textbf{DPP:} uses a drift-inspired rule combining reported pressure differences with an inverse-rate penalty to select a neighbor and an offload fraction.
\item \textbf{GATE-adapt:} integrates the graph-aware temporal encoder of~\cite{gate} into the shared satellite hybrid-action interface.
\item \textbf{ATLA-adapt:} integrates the Transformer-based encoder of~\cite{atla} into the same hybrid-action interface.
\item \textbf{Full-information reference:} uses current queue and CPU states in a sequential load-equalization heuristic, subject to the same single-neighbor offloading and ISL limits.
\end{itemize}
\rev{B1}{Local-only measures the value of cooperation; fixed-half RR tests whether observation-dependent allocation improves on a simple split.} \rev{B2}{Random feasible measures the value of informed action selection, while stale-greedy and DPP test allocation from received resource values.} The two learning adaptations provide encoder-based policy references under the shared physical evaluation settings. \rev{B3}{The full-information heuristic supplies an information-advantaged feasible reference.}

\rev{F1}{Historical backbone training budgets differ across the retained checkpoints.} \rev{F2}{Their ranking compares trained controllers, while inference ablations share initialization, fitting data, epochs, and correction limit.} \rev{F3}{Teacher training, backbone distillation, and inference fitting remain distinct preparation stages.}

Two retained preparation controls further separate backbone service from the added inference branch. The prepared backbone precedes anchored CPU inference and workload correction. \rev{F4}{The backbone-plus-correction control uses its original resource estimates with the same 0.03 correction limit.} Both share PUMA's prepared allocation backbone and evaluation dates. Their paired differences appear in Table~\ref{tab:paired}; they isolate additions to a common backbone rather than preparation-budget differences among external methods.

\subsection{Ablation Setup}
\label{sec:ablations}
Targeted interventions complement the baseline comparison by examining how individual inference components affect decisions; feature-relevance analysis~\cite{explainableV2X} provides a related perspective on interpreting learned resource allocation. The inference variants share training data, initialization protocol, 12 epochs, and correction strength. No-delay uses only the zero acquisition-delay alternative; no-history retains the latest distinct report; no-risk removes the standard-deviation term from~\eqref{eq:riskcapacity}. The temporal encoder and status-conditioned experts remain common to these variants.

Service comparisons use the reported evaluation records. Learning curves and action diagnostics are separately retained records. Service ablations quantify differences for the specified interventions, while diagnostics characterize their prediction and control channels; the two record sets are analyzed separately.

\subsection{Workload Completion}
\begin{figure}[t]
\centering\includegraphics[width=\columnwidth,trim=0bp 20bp 0bp 15bp,clip]{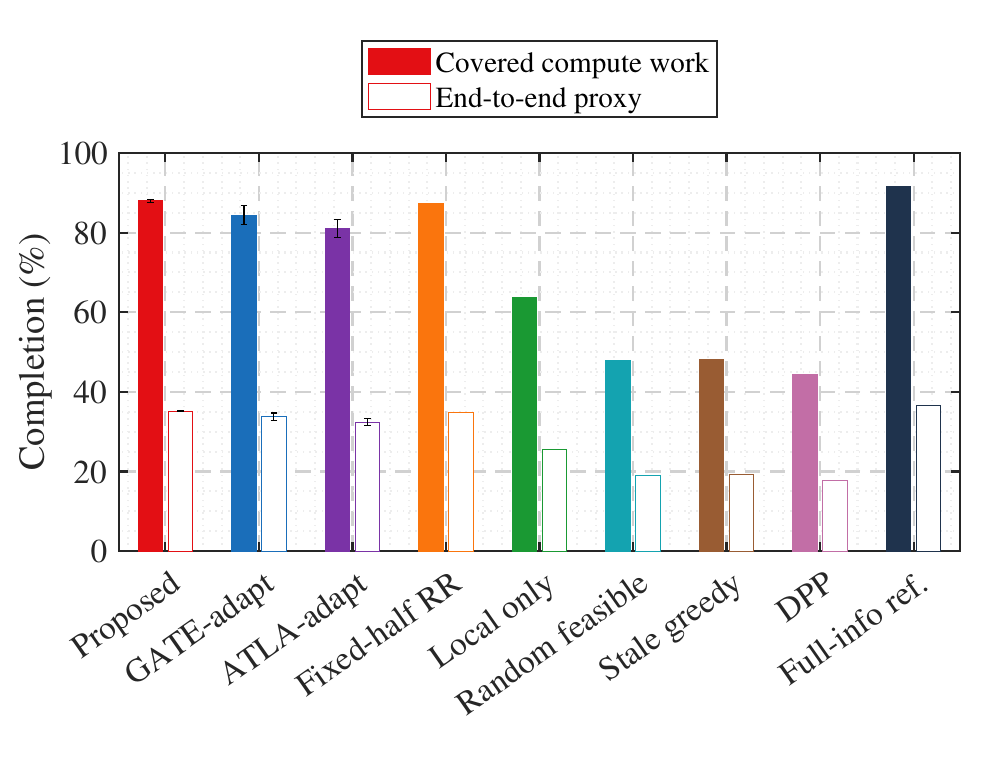}
\caption{Admitted compute-work completion across allocation methods. \rev{M2}{Filled bars normalize executed CPU cycles by admitted CPU cycles; open bars show the coverage-adjusted proxy.} Error bars show standard deviation across two training-seed means, each evaluated on eight date/scenario episodes. Rule-based methods use eight episodes. Table~\ref{tab:paired} gives conditional paired intervals for the principal comparisons.}
\label{fig:completion}
\end{figure}
Fig.~\ref{fig:completion} compares nine methods. PUMA achieves 88\% admitted compute-work completion, exceeding GATE-adapt and ATLA-adapt by 3.59 and 6.99 percentage points, respectively, or 4.26\% and 8.62\% in relative terms. Fixed-half RR reaches 87.27\%, and the full-information reference reaches 91.49\%. Local-only completes 63.82\%; random feasible, stale-greedy, and DPP achieve 47.81\%, 48.20\%, and 44.25\%.

\rev{C1}{The contrast between local-only and fixed-half RR shows the value of neighboring processors.} \rev{C2}{An outdated pressure ranking can direct stale-greedy and DPP toward processors with less available service.} \rev{C3}{PUMA uses inferred resource moments to adapt the transferred workload through a bounded partition correction, retaining the decoder's remaining resource proposal.}

\rev{F8}{PUMA adds 0.73 completion points over fixed-half RR, with a date-block interval of [0.51, 0.96] points.} Its coverage-adjusted proxy is 35.15\%. Shared access association fixes coverage across methods, so the completion difference concerns workload assignment within the reachable processing pool. The corresponding intervals against GATE-adapt and ATLA-adapt are [3.13, 4.05] and [6.26, 7.71] points under the protocol of Section~\ref{sec:protocol}.

\begin{table}[t]
\caption{Paired Differences}
\label{tab:paired}\centering\footnotesize
\setlength{\tabcolsep}{2pt}
\begin{tabular}{P{.25\columnwidth}P{.22\columnwidth}P{.22\columnwidth}P{.22\columnwidth}}
\toprule Reference & Completion (pp) & Mean delay (s) & P95 delay (s)\\\midrule
Fixed-half RR & 0.73\newline [0.51, 0.96] & $-0.03$\newline [$-0.10$, 0.04] & $-0.23$\newline [$-0.54$, 0.08]\\[3pt]
GATE-adapt & 3.59\newline [3.13, 4.05] & $-0.42$\newline [$-0.57$, $-0.27$] & $-5.84$\newline [$-11.49$, $-0.19$]\\[3pt]
ATLA-adapt & 6.99\newline [6.26, 7.71] & $-0.82$\newline [$-0.95$, $-0.69$] & $-8.42$\newline [$-11.59$, $-5.25$]\\[3pt]
Prepared backbone & 0.62\newline [0.51, 0.74] & $-0.09$\newline [$-0.12$, $-0.07$] & $-0.73$\newline [$-2.16$, 0.69]\\[3pt]
Backbone + correction & 0.21\newline [0.18, 0.25] & $-0.04$\newline [$-0.08$, $-0.01$] & $-0.25$\newline [$-0.70$, 0.20]\\
\bottomrule
\end{tabular}
\par\smallskip\raggedright Entries are PUMA minus reference, followed by conditional 95\% date-block $t$ intervals. Positive completion and negative delay favor PUMA. The P95 column compares mean episode-level P95s. The four blocks average paired scenarios and retained policy seeds; rules are not counted as independent training-seed runs.
\end{table}

\rev{F5}{The prepared backbone completes 87.38\% of admitted work at 1.63~s mean delay.} \rev{F6}{Applying the same bounded correction to its original estimates gives 87.79\% and 1.58~s.} \rev{F7}{PUMA adds 0.62 and 0.21 completion points, respectively.} These controls distinguish the prepared allocator's service level from the refinement supplied by resource inference.

\subsection{Service Delay}
\begin{figure}[t]
\centering\includegraphics[width=.94\columnwidth,trim=0bp 11bp 0bp 4bp,clip]{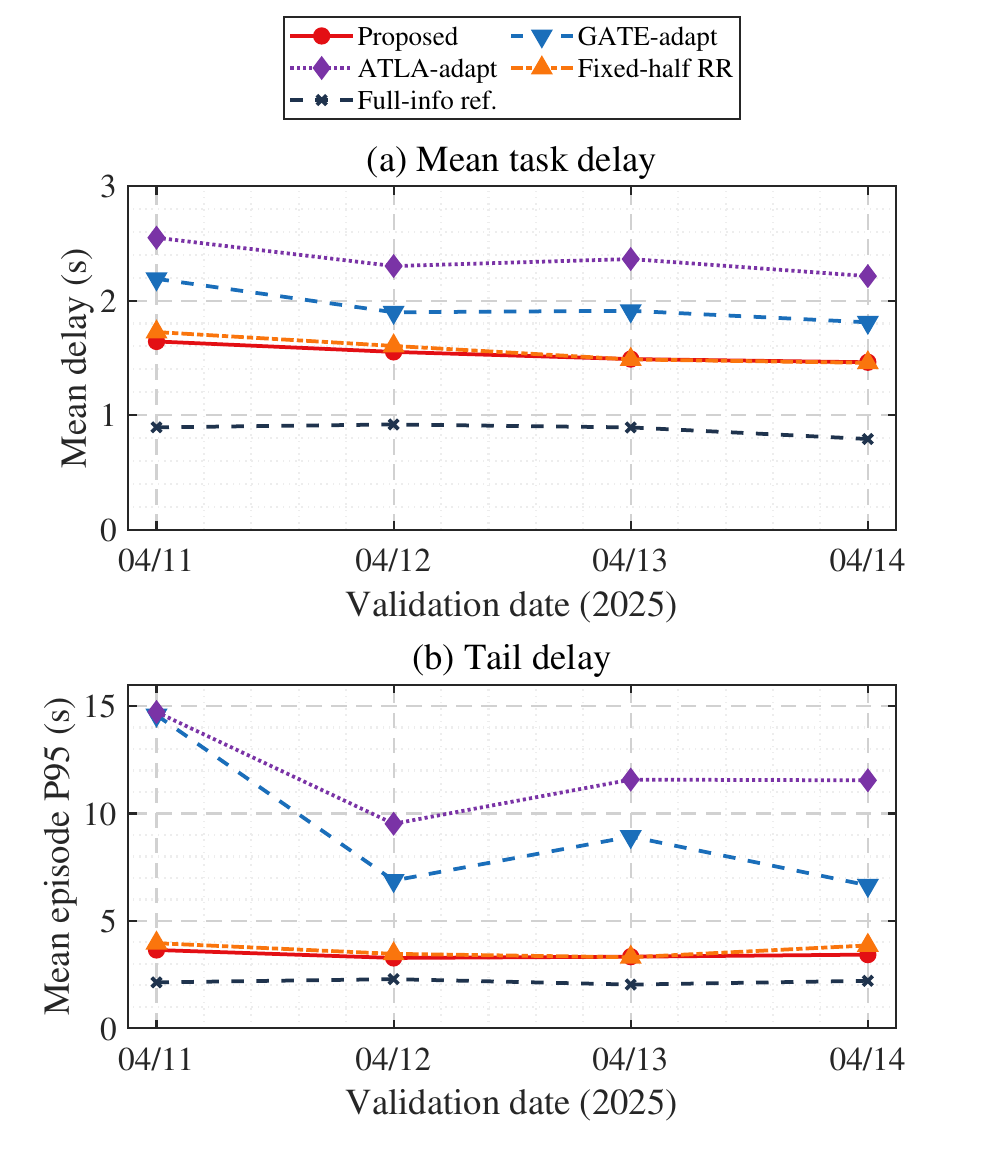}
\caption{Modeled service delay over four development dates: (a) representative-job-weighted mean delay; (b) mean episode-level P95. Date values aggregate the retained environmental scenarios and learned-policy seeds. Curves show date means under the common fluid service model; Table~\ref{tab:paired} reports the paired date-block intervals.}
\label{fig:latency}
\end{figure}
Fig.~\ref{fig:latency} reports mean and episode-P95 modeled delay. PUMA achieves 1.54~s mean delay, reducing the GATE-adapt and ATLA-adapt values by 21.28\% and 34.78\%, respectively. Its mean episode P95 is 3.42~s, compared with 9.26 and 11.84~s for the learning adaptations, corresponding to reductions of 63.02\% and 71.08\%. The date-wise curves show the same qualitative tail-delay advantage.

Fixed-half RR yields 1.57~s mean delay and 3.65~s mean episode P95; the full-information reference yields 0.87 and 2.17~s. Waiting rises nonlinearly near capacity in~\eqref{eq:wait}, making workload correction relevant to high-pressure tail assignments. \rev{F9}{Table~\ref{tab:paired} gives intervals below zero against the learned adaptations, while the intervals against fixed-half RR contain zero.} PUMA's comparison with that rule is therefore strongest in completion and normalized energy. The second-scale delay values follow the representative-job model in Section~\ref{sec:service}.

\subsection{Energy Efficiency}
\begin{figure}[t]
\centering\includegraphics[width=\columnwidth,trim=0bp 20bp 0bp 17bp,clip]{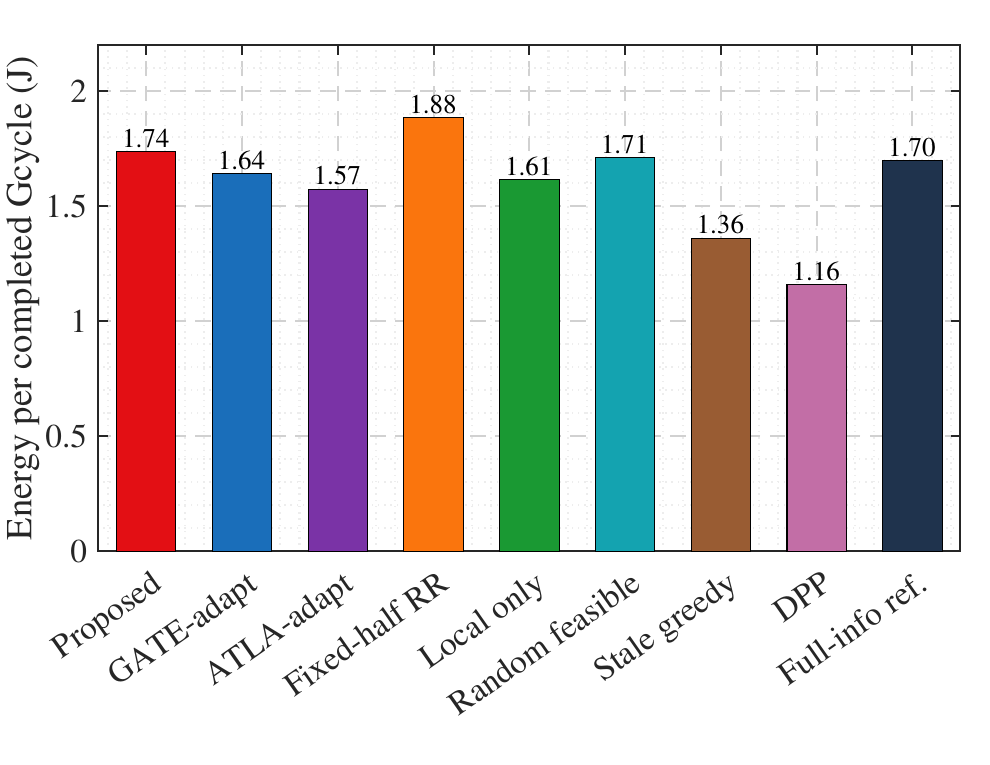}
\caption{Execution energy normalized by completed CPU work. Lower values indicate fewer joules per completed Gcycle; the completion differences in Fig.~\ref{fig:completion} must also be considered.}
\label{fig:energy}
\end{figure}
\rev{E1}{In Fig.~\ref{fig:energy}, PUMA consumes 1.74~J per completed Gcycle, compared with 1.88 for fixed-half RR, 1.64 for GATE-adapt, and 1.57 for ATLA-adapt.} \rev{E2}{Its service gains over the learning methods involve higher normalized energy, whereas it improves all three reported service--energy measures over fixed-half RR.} \rev{E3}{DPP uses 1.16~J/Gcycle at 44.25\% completion, illustrating why energy efficiency is interpreted jointly with served work.}

Offloading consumes transmission energy and changes the execution location. A capacity-sensitive split avoids transfers with little processing benefit, while larger CPU allocations can increase both completion and energy. This coupling explains the service--energy operating points.

\subsection{Module Ablations}
\begin{figure}[t]
\centering\includegraphics[width=.94\columnwidth,trim=0bp 11bp 0bp 4bp,clip]{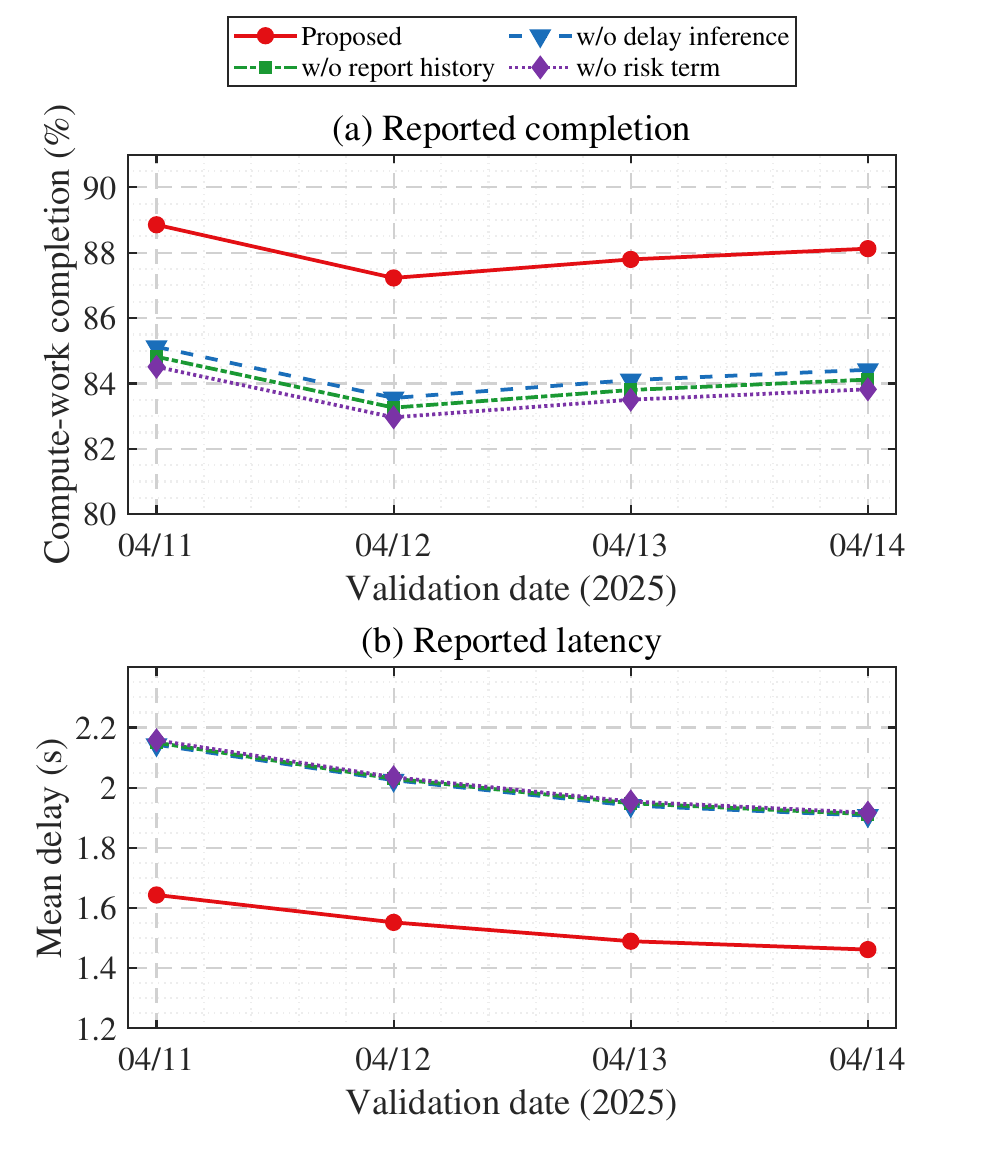}
\caption{PUMA-Former module ablations with a common correction limit: (a) admitted compute-work completion; (b) mean modeled delay. The interventions remove latent delay alternatives, older-report innovations, or the risk term in action calibration.}
\label{fig:ablation}
\end{figure}
Fig.~\ref{fig:ablation} compares matched variants at a common correction limit. No-delay, no-history, and no-risk achieve 84.30\%, 84\%, and 83.70\% completion, respectively, versus 88\% for the full model. Their mean delays are 2, 2.01, and 2.02~s, compared with 1.54~s. The reported completion decreases of 3.70--4.30 percentage points favor retaining the complete inference-to-action branch.

No-delay suppresses acquisition-age alternatives; no-history removes earlier innovations around the latest anchor; no-risk removes the direct uncertainty penalty in workload calibration. Their service losses support the complete inference-to-action branch under these interventions. The backbone and auxiliary assimilation remain common across variants.

\subsection{Inference Learning}
\begin{figure}[t]
\centering\includegraphics[width=\columnwidth,trim=0bp 0bp 0bp 16bp,clip]{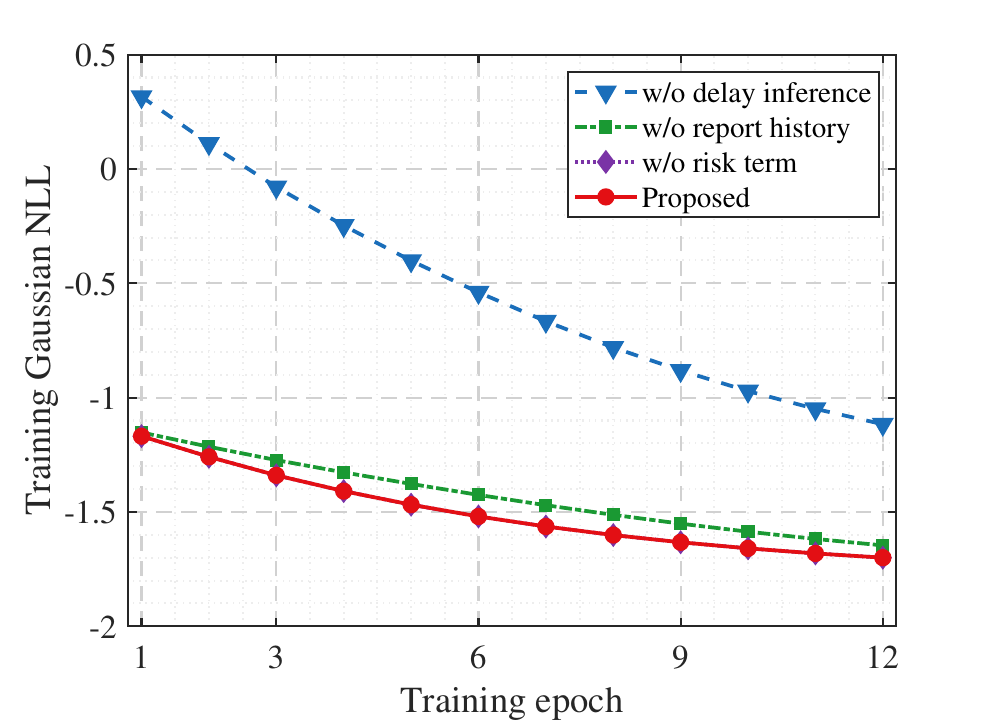}
\caption{CPU-inference learning over 12 epochs, measured by Gaussian negative log-likelihood. Full and no-risk curves coincide because their estimator training is identical; their intervention occurs in action calibration.}
\label{fig:learning}
\end{figure}
Fig.~\ref{fig:learning} shows decreasing CPU NLL. Removing latent-delay alternatives substantially increases final loss; removing earlier reports has a smaller effect. Full and no-risk coincide because estimator fitting is identical. Service metrics separately assess the decision value of this inference.

Different acquisition-delay interpretations can imply similar average CPU but different uncertainty. The identical full and no-risk curves separate estimator learning from the later use of uncertainty in control.

\rev{U1}{Table~\ref{tab:inferencechecks} complements the training curves with checks on 576 observations per seed from two training-period dates excluded from gradient fitting.} It uses the same half-uniform, half-active-node weights as~\eqref{eq:loss}. \rev{U2}{NLL includes the Gaussian constant, and coverage counts CPU targets within the pre-assimilation interval $[\mu_i-2\sigma_i,\mu_i+2\sigma_i]$.} The two seeds are averaged. \rev{U3}{These check dates also support action acceptance, placing the diagnostics within model development.}

\begin{table}[t]
\caption{Inference Checks}
\label{tab:inferencechecks}\centering\footnotesize
\begin{tabular}{P{.28\columnwidth}P{.17\columnwidth}P{.23\columnwidth}P{.21\columnwidth}}
\toprule Variant & CPU NLL & Coverage (\%) & CPU MSE $\times100$\\\midrule
Prepared backbone & $-0.46$ & 95.87 & 2.31\\
Full & $-0.78$ & 91.03 & 1.11\\
No-delay & $-0.22$ & 89.76 & 1.16\\
No-history & $-0.72$ & 88.35 & 1.11\\
\bottomrule
\end{tabular}
\par\smallskip\raggedright NLL and $\pm2\sigma$ coverage precede assimilation; MSE follows assimilation. A Gaussian $\pm2\sigma$ interval has nominal coverage 95.45\%. No-risk has the full model's inference values because it changes action correction only.
\end{table}

\rev{U4}{The full branch improves check NLL over both removals, with 91.03\% coverage against the 95.45\% Gaussian reference.} The prepared backbone has higher coverage and larger CPU error, while full and no-history have similar post-assimilation MSE. These diagnostics distinguish prediction fit, interval coverage, and decision utility. \rev{U5}{The risk term uses local approximate uncertainty through~\eqref{eq:targetrisk}--\eqref{eq:sourcerisk}; the reported coverage concerns pre-assimilation moments.}

\subsection{Action Sensitivity}
\begin{figure}[t]
\centering\includegraphics[width=.94\columnwidth,trim=0bp 34bp 0bp 15bp,clip]{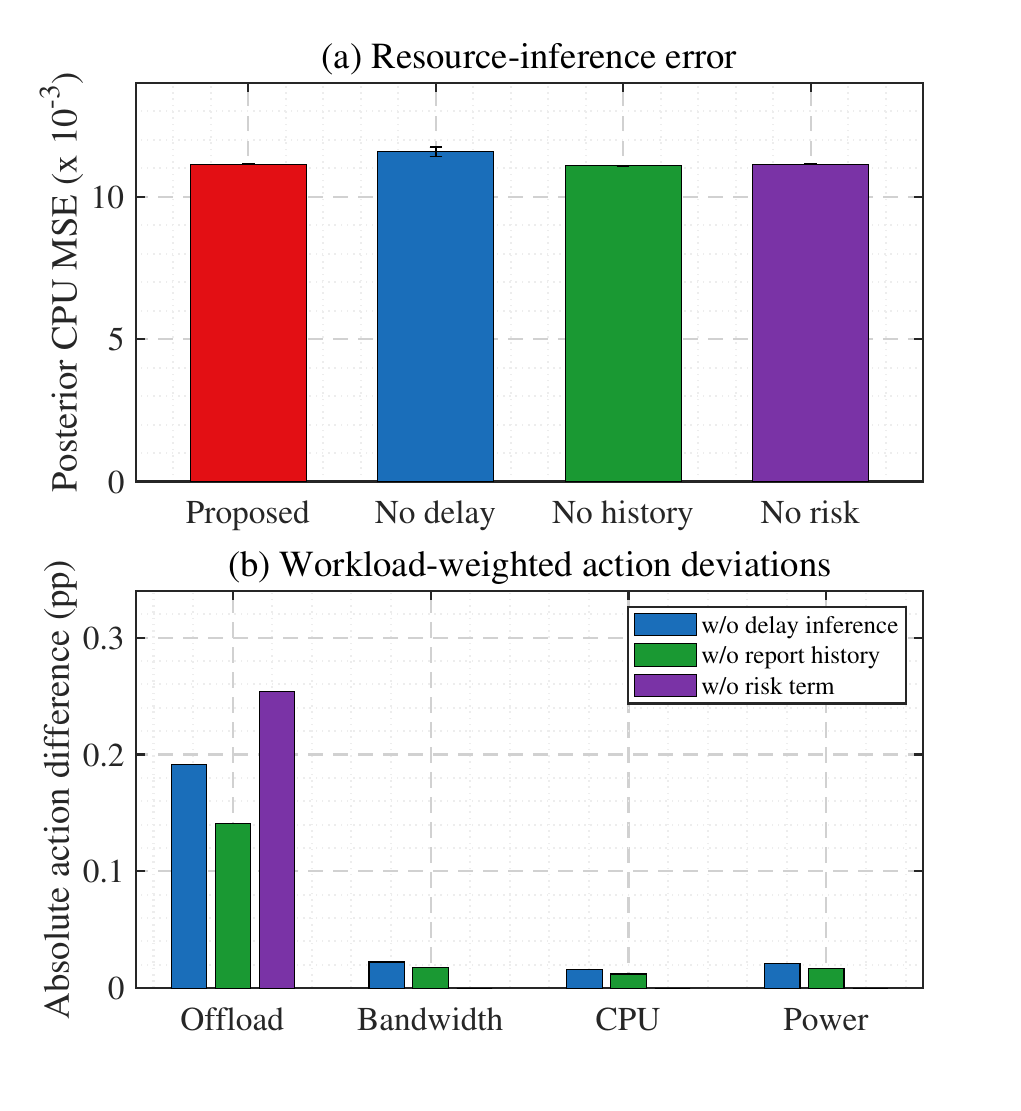}
\caption{Retained mechanism diagnostics: (a) posterior CPU error; (b) workload-weighted action deviations from the full model. These separately recorded diagnostics characterize the affected prediction and control channels.}
\label{fig:mechanism}
\end{figure}
\begin{figure}[t]
\centering\includegraphics[width=\columnwidth,trim=0bp 0bp 0bp 8bp,clip]{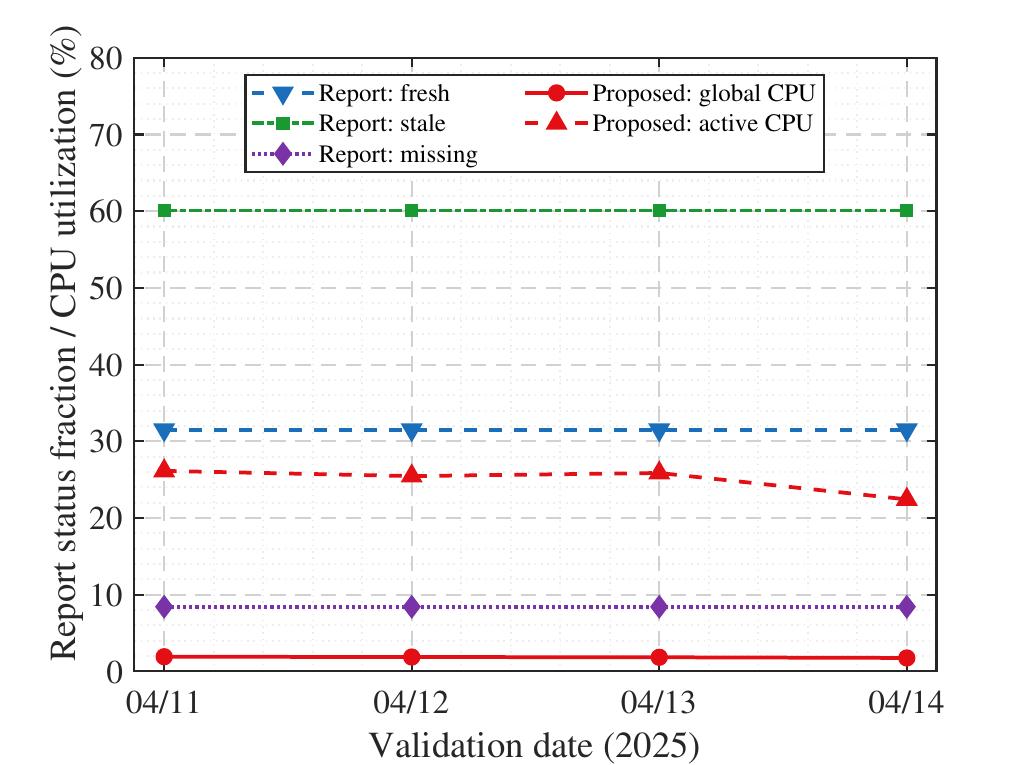}
\caption{Report freshness and CPU utilization across evaluation dates. The two red curves distinguish global utilization from utilization among active satellites; observation-status fractions describe the incomplete-information conditions.}
\label{fig:environment}
\end{figure}
In Fig.~\ref{fig:mechanism}, CPU mean-squared errors are close across variants, while workload-weighted deviations concentrate on the offload fraction, approximately 0.14--0.26 percentage points. This is the correction's target channel: near a capacity limit, a small partition adjustment can change executed work despite similar average estimation error.

These unsigned deviations locate the intervention channel;~\eqref{eq:targetrisk}--\eqref{eq:sourcerisk} give its local direction. Workload weighting emphasizes busy sources, explaining why action deviations and node-average CPU error capture different aspects of inference utility.

\subsection{Resource Conditions}
Fig.~\ref{fig:environment} shows approximately 60\% stale and 8\% missing observations, resulting jointly from report periodicity, delayed arrival, and expiry. \rev{G1}{Global CPU utilization is approximately 2\%, while active-satellite utilization is approximately 22--26\%.} \rev{G2}{The operative challenge is spatially uneven demand and restricted access to idle capacity.} PUMA uses current-resource inference to guide redistribution among reachable neighbors in this regime.

\rev{G3}{Global utilization includes all satellites; active utilization includes those carrying demand.} \rev{G4}{Idle processors outside useful access and neighbor sets cannot serve the concentrated workload, so the gains concern reachable-capacity use under Table~\ref{tab:settings}.}

\FloatBarrier
\section{Conclusion}
This study investigated cooperative computation offloading under incomplete information in MEO--LEO satellite networks, where known reachability coexists with uncertain processing capacity and competing destination loads. The resulting controller converts asynchronous report histories and auxiliary pressure into allocation-relevant resource estimates, with bounded action adjustments tied to shared workloads and local execution limits. \rev{K2}{The traffic-driven results demonstrate the value of inference-guided redistribution among reachable processors, with completion and energy advantages over fixed splitting and higher service performance at increased normalized energy relative to the retained learning adaptations.} \rev{K1}{The central implication is that resource prediction should be assessed through the workload it enables the network to execute.} Future work will integrate timestamped task arrivals and orbital contact events into service accounting, examine uncertainty after auxiliary assimilation, and jointly design indicator acquisition with offloading to capture its communication cost.

\newcommand{\BIBdecl}{\fontsize{7.7pt}{8.5pt}\selectfont}
\bibliographystyle{IEEEtran}
\bibliography{PUMA_TMC_references}

\end{document}